%% file: ms.tex
\documentclass{aastex}
\usepackage{emulateapj5}
\usepackage{amsmath}

\makeatletter

\newenvironment{inlinefigure}{%
\def\@captype{figure}%
\noindent\begin{minipage}{0.999\linewidth}\begin{center}}
{\end{center}\end{minipage}\smallskip}
\makeatother   

\journalinfo{To appear in {\it{The Astrophysical Journal Supplement Series}}}
\submitted{Received 2000 December 31; Accepted 2001 May 25}
\shortauthors{Gonzalez et al.}
\shorttitle{The Las Campanas Distant Cluster Survey}

\def\csa {counts s$^{-1}$ arcsec$^{-2}$}

\begin{document}
\title{The Las Campanas Distant Cluster Survey - The Catalog}

\author{Anthony H. Gonzalez\altaffilmark{1}, Dennis Zaritsky\altaffilmark{2},
Julianne J. Dalcanton\altaffilmark{3}, and
Amy Nelson\altaffilmark{4}}

\altaffiltext{1}{Harvard-Smithsonian Center for Astrophysics, 60 Garden Street,
Cambridge, MA 02138}
\altaffiltext{2}{Steward Observatory, University of Arizona, 933
North Cherry Avenue, Tuscon, AZ 85721}
\altaffiltext{3}{Department of Astronomy, University of Washington,
Box 351580, Seattle, WA 98195-1580}
\altaffiltext{4}{Department of Astronomy and Astrophysics, University
of California at Santa Cruz, Santa Cruz, CA 95064}

\begin{abstract}
 
We present an optically-selected catalog of 1073 galaxy cluster and
group candidates at 0.3$\la$$z$$\la$1. These candidates are drawn from
the Las Campanas Distant Clusters Survey (LCDCS), a drift-scan imaging
survey of a 130 square degree strip of the southern sky. To construct
this catalog we utilize a novel detection process in which clusters
are detected as positive surface brightness fluctuations in the
background sky. This approach permits us to find clusters with
significantly shallower data than other matched-filter methods that
are based upon number counts of resolved galaxies.  Selection criteria
for the survey are fully automated so that this sample constitutes a
well-defined, homogeneous sample that can be used to address issues of
cluster evolution and cosmology. Estimated redshifts are derived for
the entire sample, and an observed correlation between surface
brightness and velocity dispersion, $\sigma$, is used to estimate the
limiting velocity dispersion of the survey as a function of
redshift. We find a net surface density of 15.5 candidates per square
degree at $z_{est}\ge$0.3, with a false-detection rate of
$\sim$30\%. At $z$$\sim$0.3 we probe down to the level of poor groups
while by $z$$\sim$0.8 we detect only the most massive systems
($\sigma$$\ga$1000 km s$^{-1}$).  We also present a supplemental
catalog of 112 candidates that fail one or more of the automated
selection criteria, but appear from visual inspection to be {\it{bona
fide}} clusters.

\end{abstract}

\section{Introduction}

 In the quest to determine the parameters describing cosmological
models and discriminate between the various models, galaxy clusters
constitute a uniquely powerful class of objects. In contrast to the
galaxy distribution, the cluster distribution remains closely coupled
to the initial power spectrum, probing scales where the mass
distribution is still governed by linear dynamics. Consequently, it is
relatively simple to extract information about cosmological parameters
from properties of the cluster population.  Properties such as the
cluster abundance and spatial correlation length are strongly
dependent upon $\Omega_0$, but are insensitive to
$\Omega_\Lambda$. Cluster-based constraints therefore complement
cosmic microwave background (CMB) and high-redshift supernovae
constraints, which are sensitive to $\Omega_0$+$\Omega_\Lambda$ and
$\Omega_0$-$\Omega_\Lambda$, respectively.

Unfortunately, the potential for strong, cluster-based constraints
remains largely unrealized. A major limitation has been the dearth of
known clusters at $z$$>$0.5, because it is at these redshifts that
model predictions strongly diverge \citep{whi93}.  The two most common
techniques for finding distant clusters are optical searches for
projected galaxy overdensities and X-ray searches for extended thermal
bremstraahlung, but the effectiveness of both approaches is currently
limited at $z$$>$0.5. In the optical, detection of projected
overdensities requires deep imaging.  As a result, only relatively
small areas have been thus surveyed, with the largest published survey
of this kind being the I-band ESO Imaging Survey
\citep[EIS;][]{sco99}, which covers 17 square degrees.  Further,
aggregates of cluster galaxies are dominated by faint field galaxies,
and so detection of true overdensities is difficult. Detection
efficiency can be improved by including color information in the
search for aggregates, but requires either a reduction in survey area
or corresponding increase in telescope time.  In the X-ray, the
problem is not angular coverage but rather detector sensitivity. The
most recent generation of orbital telescopes had insufficient
sensitivity to detect all but the most luminous high-redshift
clusters. Only six clusters at $z$$>$0.5 were discovered in the
Einstein Medium Sensitivity Survey (EMSS; \citealt{hen92};
\citealt{gio94}), and the largest published X-ray sample for this
redshift regime is a set of 24 clusters detected as part of a cluster
survey by \citet{vik98} using archival ROSAT PSPC data.

With the Las Campanas Distant Cluster Survey (LCDCS) we generate a
catalog that incorporates some of the most desireable elements of each
of the two traditional approaches.  By employing a novel technique for
identifying clusters, we are able to survey an effective angular area
that is a factor of five larger than traditional optical surveys,
while probing to higher redshift and lower mass limits than the
existing X-ray surveys. Further, we examine a sample of known clusters
and utilize the properties of these systems, in conjunction with
follow-up imaging of a subset of LCDCS clusters, to calibrate methods
of estimating the redshift and velocity dispersion, $\sigma$, for all
candidates. Estimating $\sigma$ is critical, as the principal
advantage of X-ray surveys over optical surveys has been that X-ray
luminosity, $L_X$, is much more strongly correlated with mass than
optical richness. The survey concept is explained in \S
\ref{sec-concept}, details of the reduction procedure and cluster
identification are described in \S \ref{sec-redux}-\ref{sec-id}, and
the catalog is presented in \S \ref{sec-catalog}.  The properties of
the sample are discussed in \S \ref{sec-properties}, and \S
\ref{sec-discussion} contains a summary and discussion of the desired
properties of future surveys utilizing this approach.

\section{Survey Concept}
\label{sec-concept}

The basic idea driving the Las Campanas Distant Cluster Survey is that
clusters can be detected as regions of excess surface brightness
relative to the background sky. This hypothesis, first suggested by
\citet{dav94} and developed in detail by \citet{dal96}, posits that
although few individual cluster galaxies may be detectable, the
integrated signal from the undetected galaxies can be sufficiently
large for detection of the cluster. Subsequently, consideration of the
luminosity budgets of local clusters (\citealt{uso91,sch94,gon2000})
has led to the realization that this signal is further augmented by a
significant contribution from the halo of the brightest cluster
galaxy. To maximize the contrast between a cluster and the background,
the image should be smoothed on a scale comparable to the core size of
the cluster, thus reducing the Poisson noise.  Pilot work utilizing
drift-scan data from the Palomar 5m demonstrated the feasibility of
this approach (Dalcanton 1995; Zaritsky et al. 1997; data described in
Dalcanton et al. 1997), laying the groundwork for the LCDCS.

   Our approach offers several key advantages relative to other recent
optical surveys, which all rely upon identification of overdensities
in the projected galaxy number density to detect clusters. Most of
these surveys, such as the Palomar Distant Cluster Survey
\citep[PDCS;]{pos96}, employ a weighted filter designed to match the
expected cluster luminosity function and radial profile at a given
redshift.\footnote{Other examples of the use of matched filters with
projected number counts can be found in \citet{kaw98}, \citet{ols99},
and \citet{kep99}. The latter group expands upon previous work by
employing an adaptive matched filter capable of incorporating
photometric and spectroscopic redshift information.}  One key
advantage of our approach is that we require much shallower imaging
than previous optical surveys because our detection technique does not
require that we resolve individual cluster galaxies.  For the LCDCS,
we are able to detect clusters out to z$\sim$1 with an effective
exposure time of 194s on a 1m telescope. This approach permits a large
area to be surveyed, which is necessary for detection of the richest,
rarest systems.  A second advantage is that, because this method is
sensitive only to dense cluster cores, cluster detections have a small
cross-section ($\sim20\arcsec$, or $\sim$100 $h^{-1}$ kpc at $z$=1).
Consequently, detections due to superposition of poor systems or the
presence of wall-like structures are rare.  Further, this method is
less dependent upon the cluster luminosity function than surveys that
depend upon number counts (e.g. a cluster can be detected via the BCG
halo in the absence of other bright cluster galaxies) and makes no
assumption about galaxy colors or the presence of a well-defined red
sequence.  As a result, comparison of this type of survey with more
traditional optical catalogs should be quite productive for better
defining the selection biases of both methods.  The key disadvantage
of this approach is that a variety of astrophysical phenomena are
capable of inducing surface brightness excessess (most notably
galactic cirrus, low surface brightness galaxies and tidal tails), and
techniques must be developed to minimize contamination of the catalog
by these sources. The methods employed for minimizing contamination in
the LCDCS are discussed in \S \ref{sec-id}.

\begin{inlinefigure}
\plotone{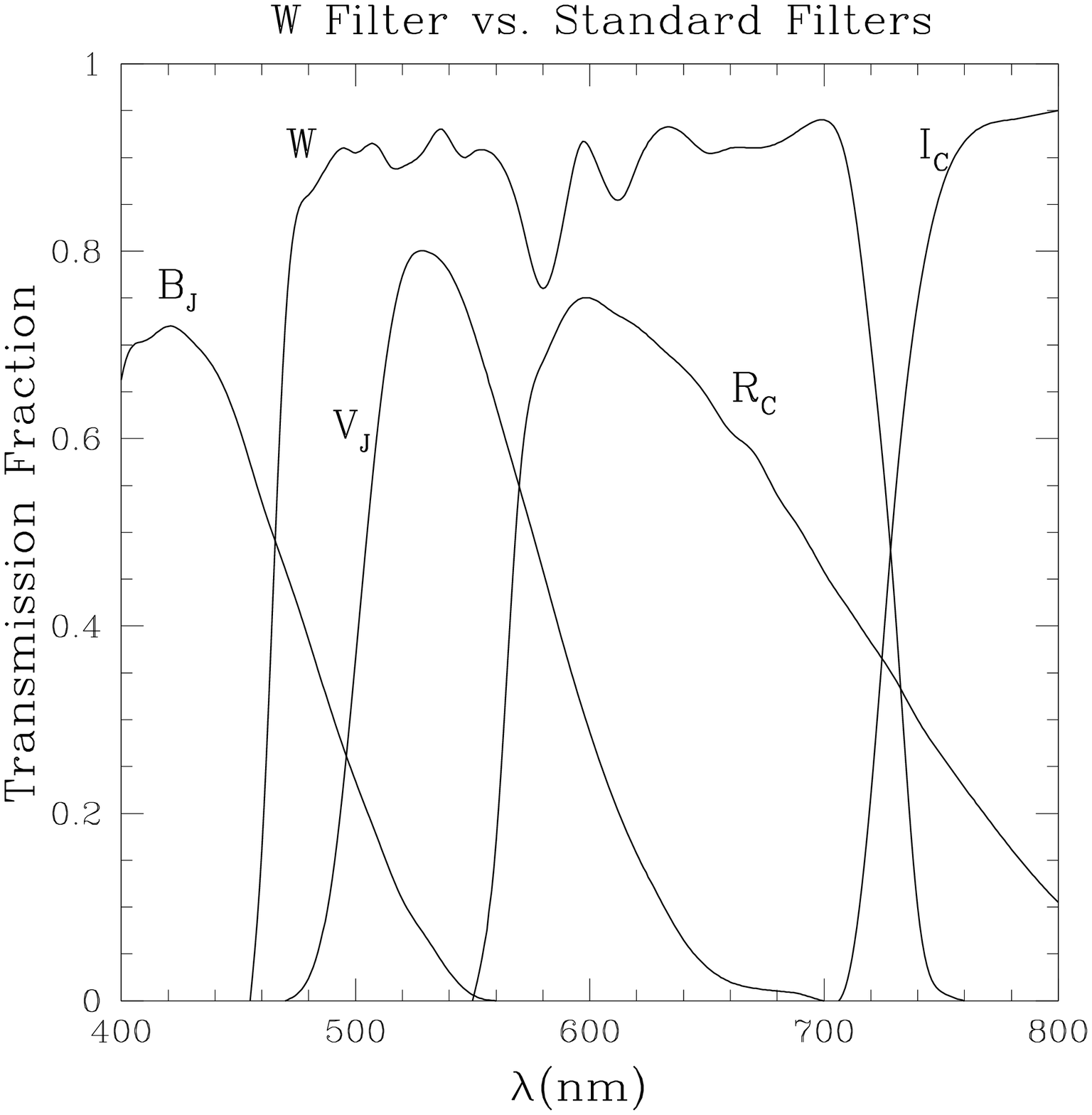}
\figcaption{Comparison of the broad $W$ filter utilized in
this work to standard Johnson ($BV$) and Cousins ($RI$) filters.  The
red cutoff is designed to avoid night sky lines, while maximizing
incident flux. The blue cutoff is designed to avoid large atmospheric
refraction.\label{fig:filter}}
\end{inlinefigure}

\section{The Data}

The survey data were obtained in March 1995 under photometric
conditions using the Las Campanas 1m telescope, the Great Circle
Camera \citep{zar96}, and the Tek\#5 CCD.  We employed a custom,
wide-band filter (hereafter designated $W$) designed to maximize the
incident flux while avoiding strong atmospheric emission lines in the
red and atmospheric refraction in the blue. The wavelength coverage,
which roughly extends from $B$ to $I$, is shown in Figure
\ref{fig:filter}.  Individual drift scans are 2048$\times$20000 pixels
with a plate scale of $0\,\farcs697$ pixel$^{-1}$ and an effective
exposure time of 97s. The data consist of 198 contiguous, overlapping
scans that collectively cover 160 square degrees of the southern sky.
The geometry is such that nine scans are obtained at a given right
ascension.  Each of these nine scans is shifted in declination by half
the width of the CCD from the previous scan. With this approach, we
cover a strip extending 85$\degr$ in right ascension
(10h$<$$\alpha$$<$15h39m) with a width of 1.8$\degr$ in declination.
For the cluster survey, we use the central 1.5$\degr$ in declination
($-13\degr$$<$$\delta$$<$$-11\degr 30\arcmin$), for which every
location is imaged twice. The net exposure time of this region is 194
seconds.

Magnitudes are calibrated using standard fields from \citet{lan92}. To
permit calibration of the $W$ filter, images of the standard stars
were obtained in Cousins $R$- and $I$- as well as in the $W$-band. We use
this data to define the zeropoint of the $W$ filter on the Vega system (i.e.
$W$=0 for Vega).
\noindent For
reference, the galaxy density and galaxy to star ratio are plotted in
Figure \ref{fig:stargal} as a function of $W$-band magnitude.
\begin{inlinefigure}
\plotone{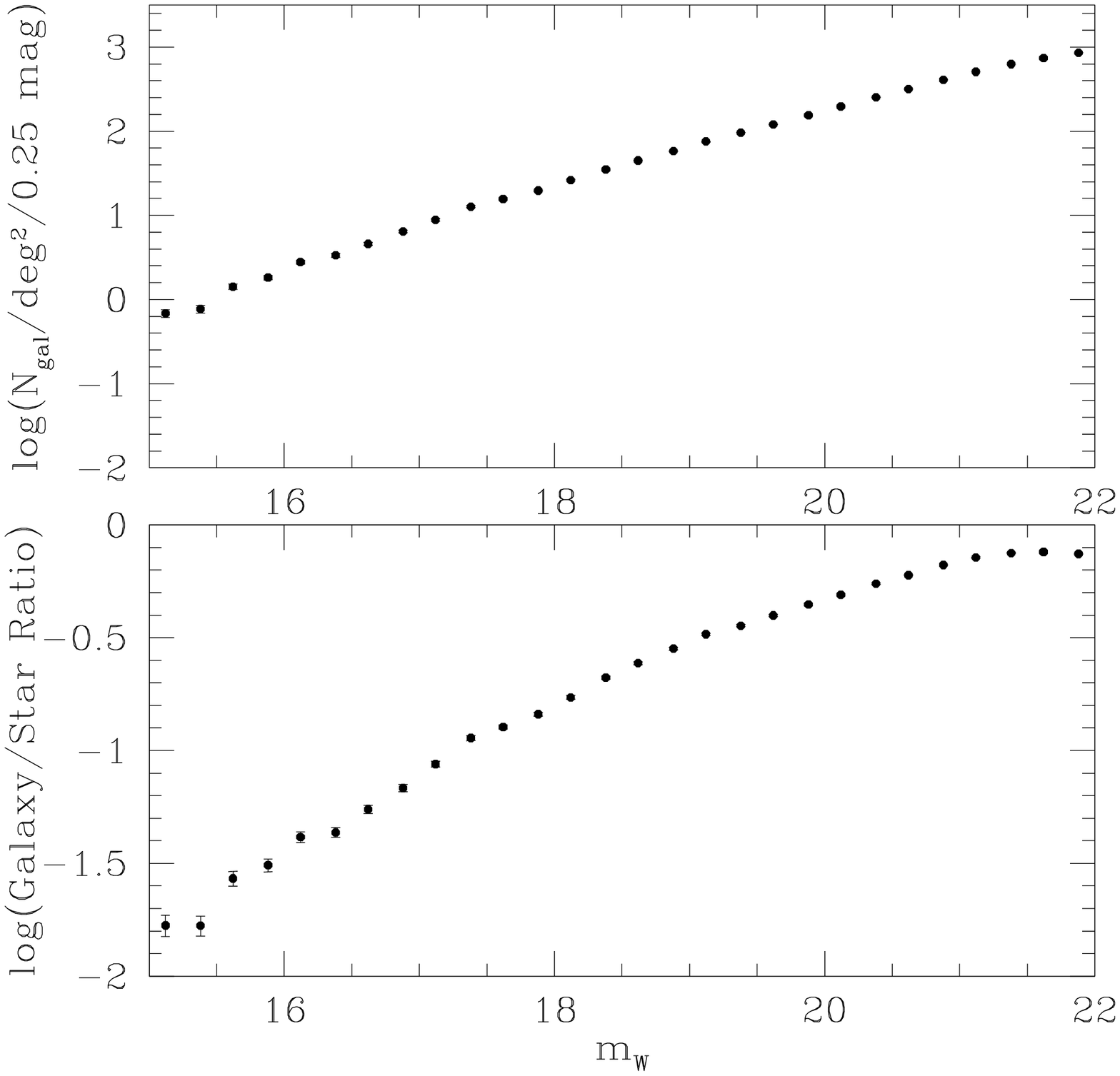}
\figcaption{Projected galaxy density and ratio of galaxies to
stars as a function of m$_W$.\label{fig:stargal}}
\end{inlinefigure}
%

\section{Reductions}
\label{sec-redux}
 
\begin{figure*}
\epsscale{0.6}
\plotone{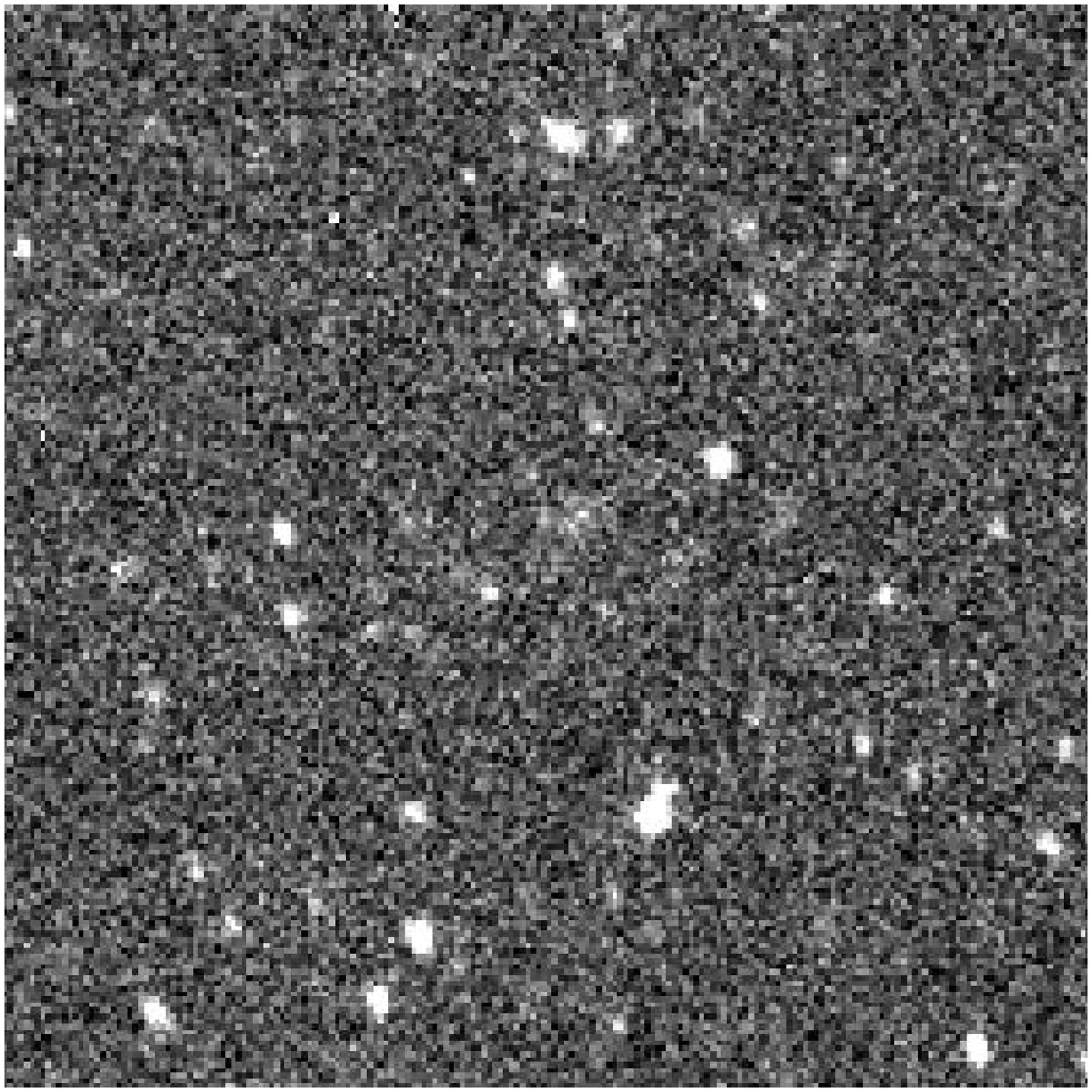} \plotone{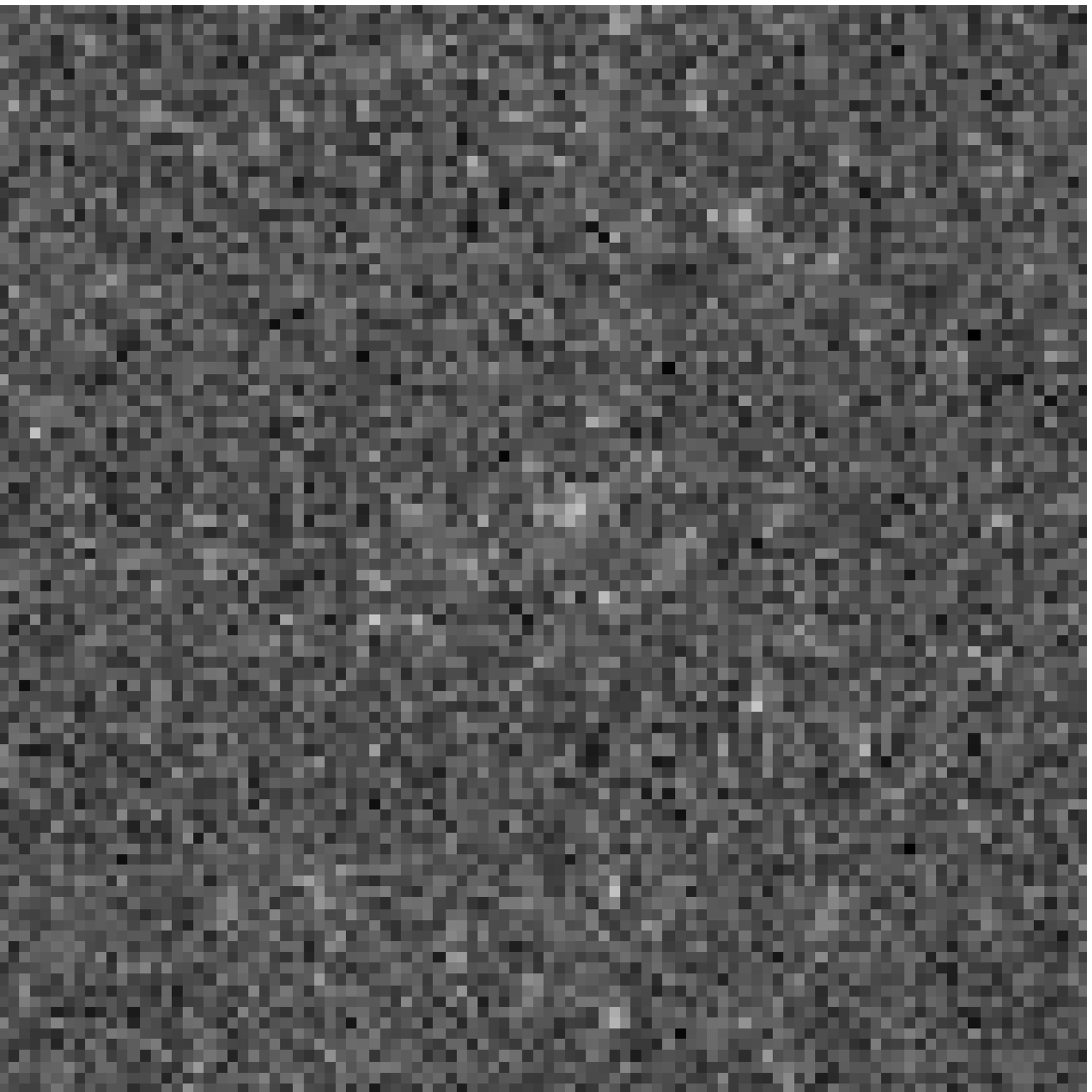} \plotone{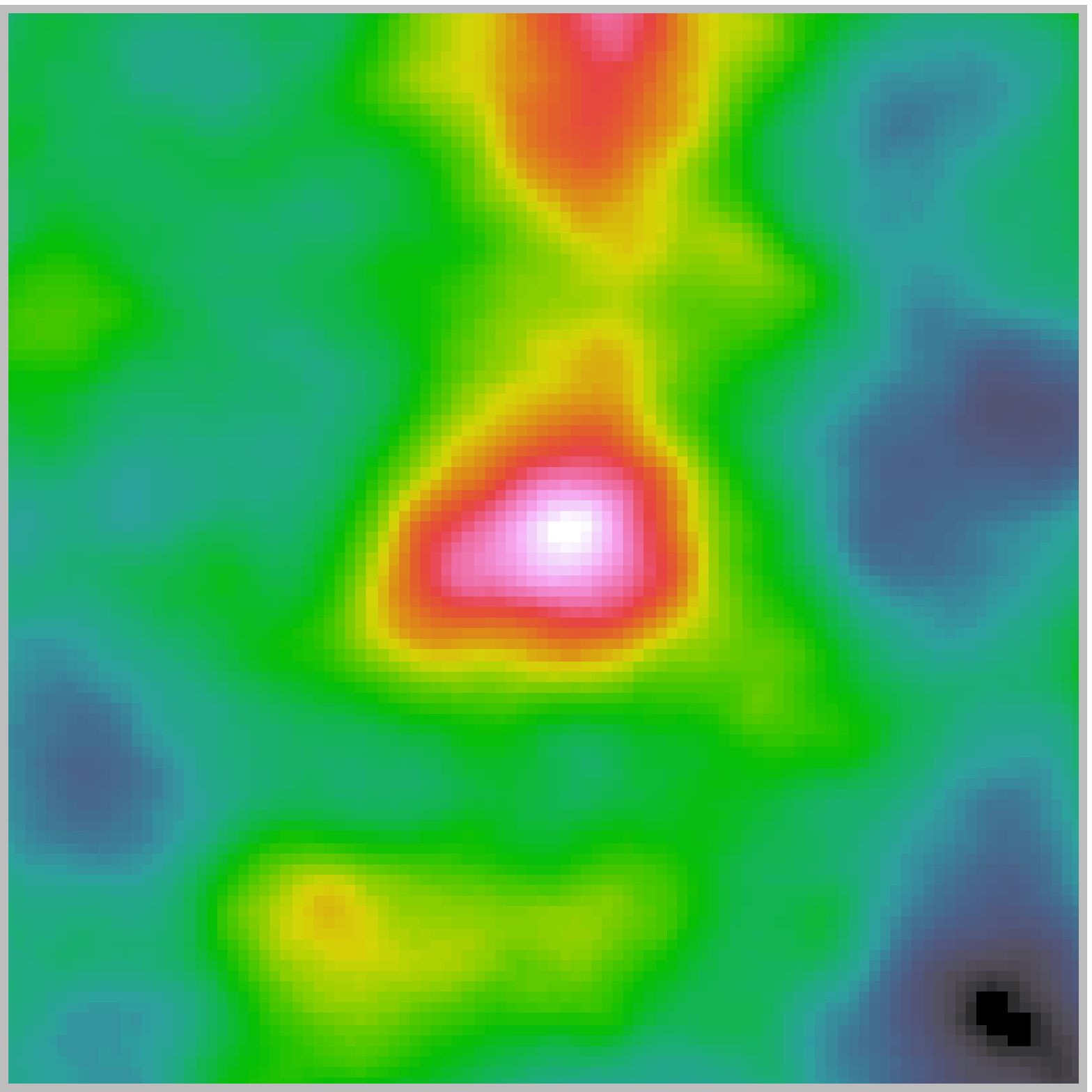}\\
\figcaption{ Three panel image illustrating the
reduction process for a $140\arcmin\times140\arcmin$ region, centered
on a cluster with a spectroscopic redshift $z$=0.80. The left panel is
the bias-subtracted, flatfielded drift-scan data. The middle panel
shows an intermediate stage in the reduction process in which detected
objects have been replaced with locally drawn, random sky pixels. This
image is then convolved with an exponential smoothing kernel of scale
length 10$\arcsec$, resulting in the image shown in the panel on the
right. Sextractor is utilized to detect positive fluctuations in the
smoothed data.
\epsscale{1.0}
\label{fig:process}}
\end{figure*}

  An overview of the reduction procedure employed is as follows.
First, individual scans are bias-subtracted and flatfielded. Next, an
image registration routine is used that calculates the image center
and curvature of each scan relative to a cartesian system defined with
axes ($\alpha$,$\delta$). We then use FOCAS to detect all objects in
the scans and ``clean" detected objects down to a limiting aperture
magnitude $m_{ap}$=22.8 mag. Bright stars and galaxies are masked at
this stage and the image is binned by a factor of two.  After masking,
we subtract temporal sky fluctuations from the individual scans and
perform a second stage of flatfielding.  Next, the results of the
image registration stage are used to geometrically transform the
images into the cartesian coordinate system.  Remaining large-scale
brightness fluctuations are subtracted via boxcar smoothing, and sets
of images with the same right ascension are combined into large
mosaics.  Finally, we convolve the mosaics with an exponential
smoothing kernel with a scale length of 10$\arcsec$ to enhance
cluster-size surface brightness features. SExtractor is used to detect
fluctuations in the smoothed mosaics and an automated classification
routine is employed to determine which fluctuations are induced by
clusters. An illustration of this procedure can be seen in Figure
\ref{fig:process}, and the various aspects of this procedure are
described in detail below.

\subsection{Bias Subtraction and Flatfielding}  

Bias subtraction is performed for the drift-scan data using overscan
regions in both dimensions.  The overscan region perpendicular to the
readout direction is 30 pixels wide; the overscan region parallel to
the direction of readout is 98 pixels wide for half of the scans, and
10 pixels wide for the rest. To model the bias, we average across the
width of the overscan region, and then apply a symmetric
Savitsky-Golay smoothing filter (60 pixels in width) along the length
of the overscan region to reduce pixel-to-pixel noise \citep{pre92}.
Rms variations in the subtracted bias are $\sim$0.1\%. High precision
in this step is important, as residual variations can induce spurious
detections. Fortunately, removal of the bias is augmented by
subsequent reduction stages.  For example, some of the residual
variation along the readout direction will be removed during sky
subtraction, and in both directions residual variation is damped by
$\sqrt{2}$ when the images are combined.

Accurate flatfielding of the data is also critical. The reduced data
must have residual flatness variation of less than 0.2\% ($\mu_W\sim$
28.3 mag arcsec$^{-2}$), or else these variations will be the dominant
source of noise limiting cluster detection.  A key property of
drift-scan data is its intrinsic uniformity.  With drift scans,
pixel-to-pixel variation is minimized because data are clocked across
the chip, so sensitivity variations are a concern only perpendicular
to the readout direction (at a level $\sim$ 2\% in our raw
data). Consequently only a one-dimensional flatfield is required, for
which the Poisson noise can be reduced significantly relative to a
two-dimensional flatfield by averaging along the direction of readout.
To reduce the flatness variation to the desired level, we apply two
stages of flatfielding. In both stages we use a set of 4 or 5 scans to
construct a sigma-clipped, median averaged flatfield. (For a given
right ascension, typically four of the scans are taken consecutively
one night, and the other five are taken consecutively on a different
night.)  Each scan is 20,000 pixels in length and the typical sky
level in $W$ is $\sim$100 counts, so the associated Poisson noise is
0.035\% per column. The first stage of flatfielding immediately
follows bias subtraction and reduces sensitivity variation from 2\% to
0.4\%. Subsequent to this stage, we use the IRAF routine {\it{fixpix}}
to interpolate across the three bad columns in our drift scans (see
Figure \ref{fig:flatfield}).  The second flatfielding stage is
designed to remove the 0.4\% residual variation.  To eliminate
contamination from resolved objects, which may lead to correlated
variations across columns, this second stage is run after all detected
objects have been replaced with local sky pixels (see \S
\ref{subsec-cleaning}), and after temporal fluctuations in the sky
level have been subtracted along the direction of readout (see
\S\ref{subsec-sky}).  Figure \ref{fig:flatfield} shows a typical pair
of flatfields resulting from these two stages of flatfielding.
Subsequent to this final flatfielding all scans are flat to
$\la$0.2\%, with an rms residual variation of $\sim$0.1\%.

\subsection{Object Removal and Masking}
\label{subsec-cleaning}
Next, we remove all stars and galaxies that are individually
detectable in the images using a modified version of the Faint Object
Classification and Analysis System (FOCAS) (Jarvis \& Tyson 1981;
Valdes 1993; modifications described in Dalcanton et al. 1997).  The
basic strategy is to use FOCAS to detect all objects in the scans and
then clean out objects down to a limiting aperture magnitude.  Using
an aperture of radius 5 pixels, we choose a limiting aperture
magnitude m$_W$=22.8, which roughly corresponds to a 4-$\sigma$
detection in an individual scan.

\begin{inlinefigure}
\epsscale{1}
\plotone{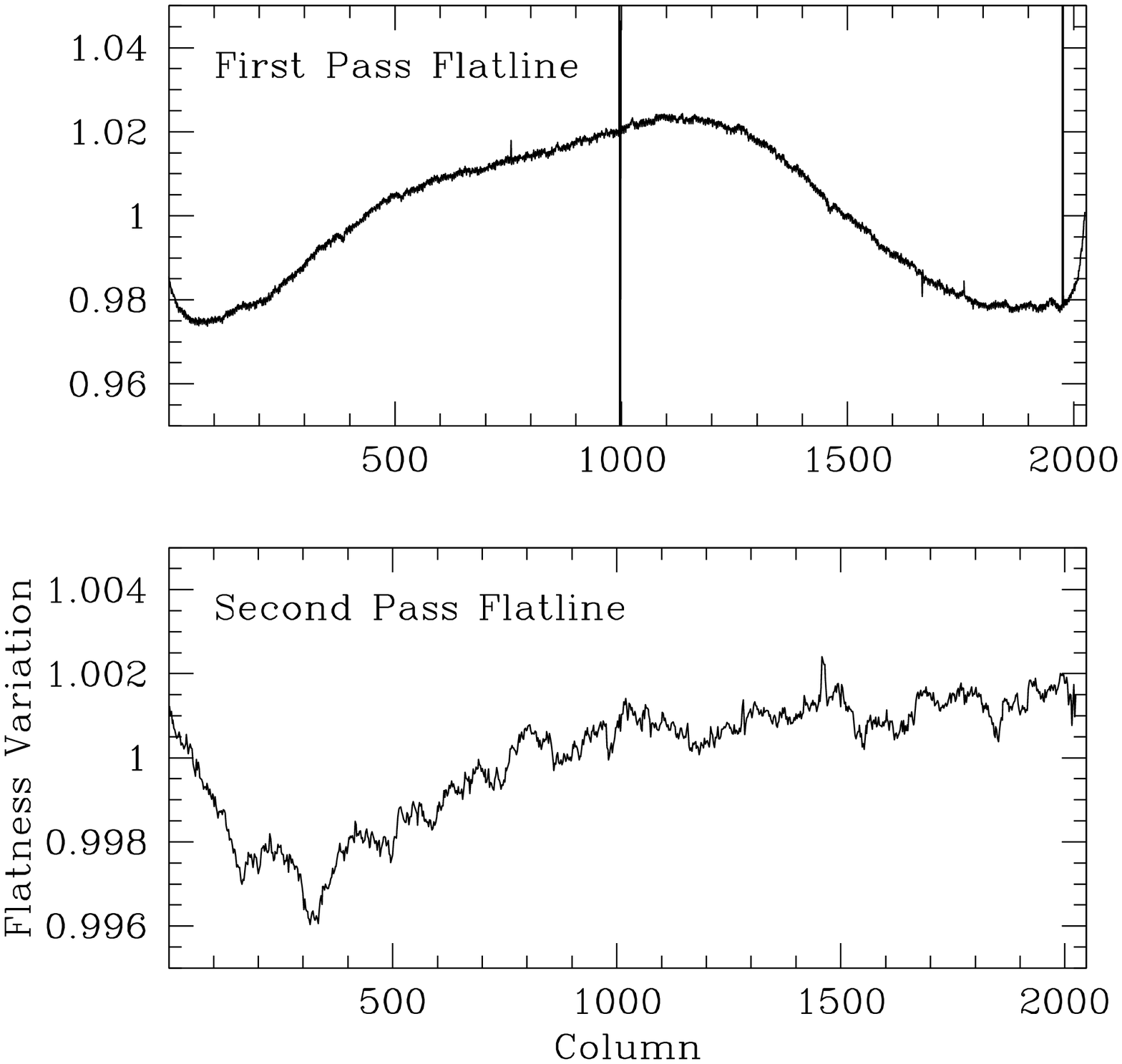}
\figcaption{ Typical flatfields
generated in reducing the scans. The upper panel shows the 1-D flatfield 
applied to the raw data, which corrects initial variations of order
2\%. The sharp vertical features in this panel correspond to bad columns.
The lower panel shows the second pass flatfield which is
applied to scans after sky subtraction and cleaning. Subsequent to this
stage, residual variations are at the level of 0.1\%. 
\label{fig:flatfield}}
\end{inlinefigure}
	
One inherent obstacle in using FOCAS for detection is that drift scans
have a time-variable component to the sky level. This variability in
turn means that the rms of the sky will vary through the image.
Because FOCAS uses a fixed sigma threshold throughout the scan, the
sky variation results in variable depth for the FOCAS catalog.  The
solution we employ is to calculate and input to FOCAS the sigma
corresponding to the lowest sky level in the scan. The catalog then
goes too deep in some locations, but is deep enough everywhere to
uniformly implement the desired aperture magnitude limit for cleaning,
and yields uniform object removal throughout the image.  The actual
cleaning is accomplished with a modified version of the FOCAS routine
CLEAN, which replaces object 
\begin{inlinefigure}
\plotone{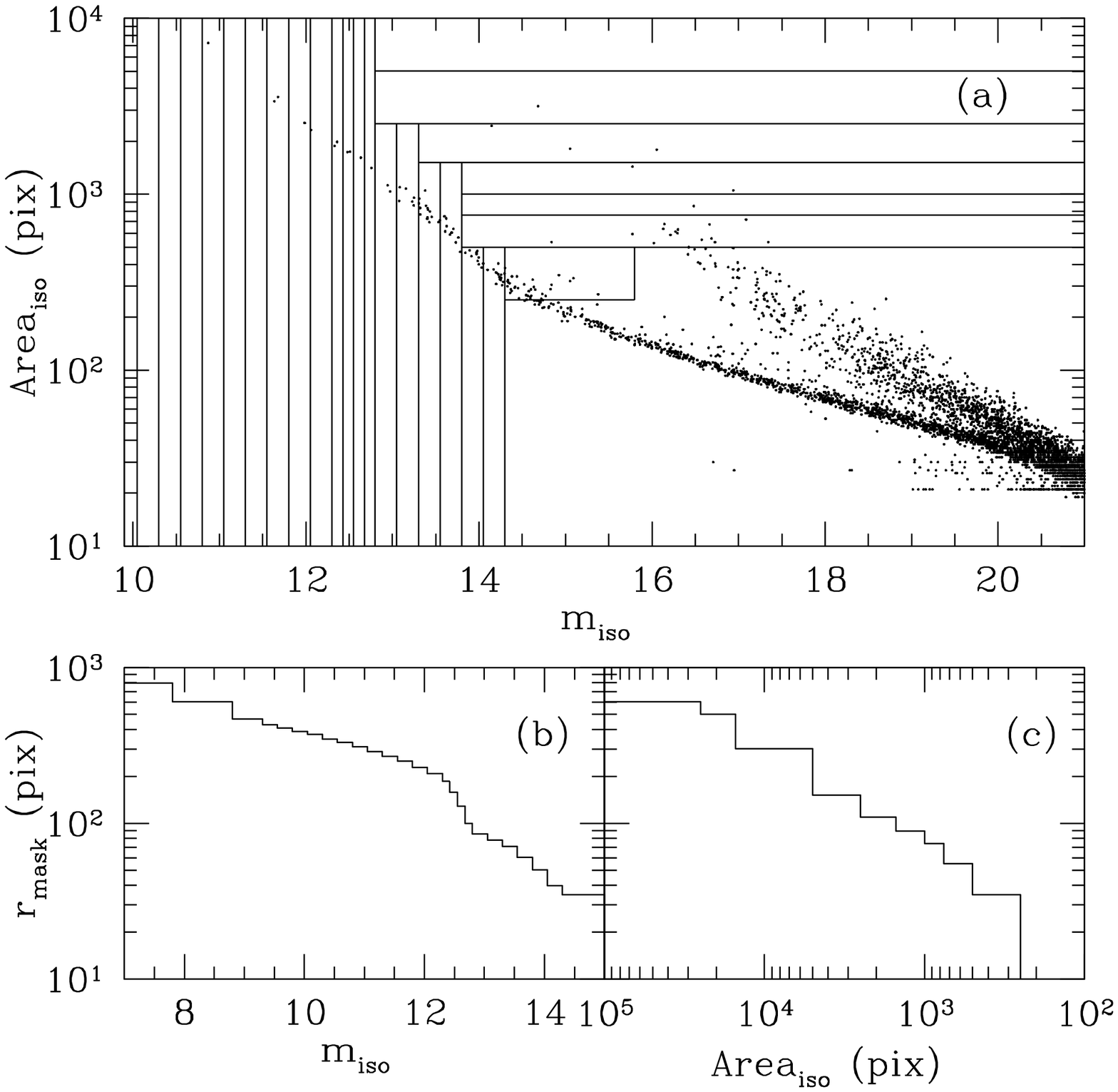}
\figcaption{ (a) Mask size as a function of
FOCAS isophotal magnitude and isophotal area. Points indicate a
representative subset of objects detected in the survey, with the
lower branch containing stars and the upper branch containing
galaxies.  Solid lines demarcate changes in mask size, with no masking
applied to objects in the lower right region of the plot. For stars,
isophotal magnitude is the primary factor driving the mask size; for
galaxies, isophotal area is the more important factor. An inflection
occurs in the stellar branch at m$\sim$15 and corresponds to the
magnitude above which stars are saturated in this data set. (b)
Stellar mask size as a function of isophotal magnitude. (c) Galaxy
mask size as a function of isophotal area.
\label{fig:masking}}
\end{inlinefigure}

\noindent pixels with local sky pixels. When
cleaning, the replacement region corresponds to the isophotal area of
the detection, plus a growth region, the width of which is a function
of the 5$\arcsec$ aperture magnitude of the object.  For objects with
$m_{ap}<16.8$, the width of the growth region is 8 pixels; for objects
just above the cleaning threshold the width of the growth region is 2
pixels.  There are two notable cases in which this cleaning proves
insufficient, however.  Saturated stars and large galaxies both leave
residual flux in 
\begin{inlinefigure}
\plotone{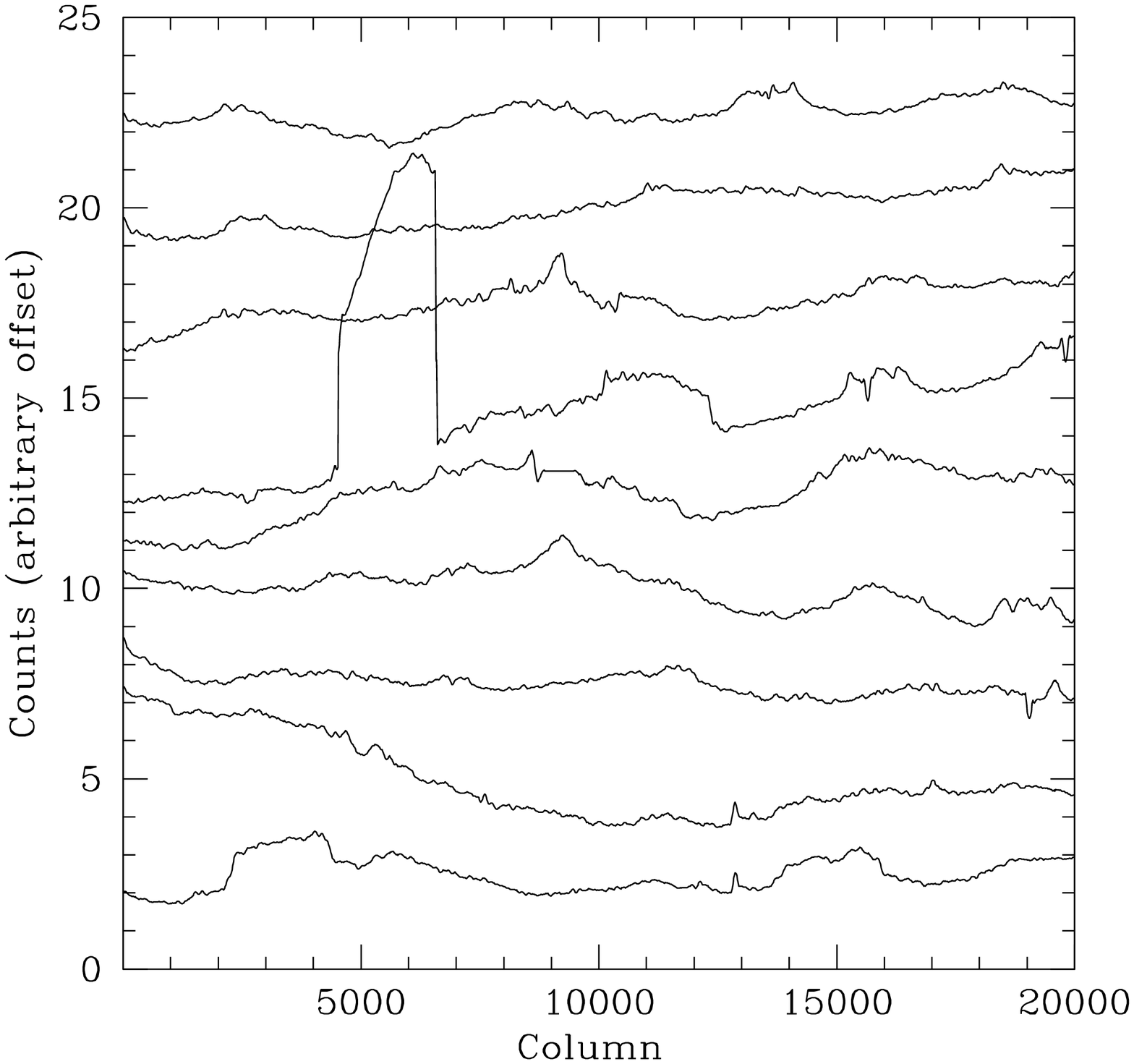}
\figcaption[f6.eps]{ Sky fits for a set of nine scans
at the same right ascension. The mean level of a given fit is
arbitrary as these have been renormalized and plotted in order of
decreasing declination, with a mean offset of 2.5 counts between each
scan. This figure illustrates the amplitude of the sky fluctuations,
and also the need for a discontinuous component to the sky level in
some cases. \label{fig:skyfit}}
\end{inlinefigure}

\noindent the image beyond the cleaning radius, and
consequently are masked.

To mask, we identify the objects for which cleaning is insufficient.
Figure \ref{fig:masking}$a$, which shows isophotal area as a function
of isophotal magnitude, illustrates how we perform this
identification. Objects on the upper branch in this figure are
galaxies; those on the lower branch are stars.  For large galaxies, we
use the isophotal area to determine the size of the mask.  For the
saturated stars (m$_W\la$15), we instead use the isophotal magnitude.
The radii of the circular masks as a function of isophotal area and
magnitude are shown in Figures \ref{fig:masking}$b$ and
\ref{fig:masking}$c$. For both stars and galaxies we take a
conservative approach and apply large masks, with the typical mask
radius for galaxies being $\sim5$ times the isophotal radius.  In
addition to bright stars and galaxies, scan edges and linear features
in the data such as bleed trails are also masked.  The net effect of
all masking is a 28\% reduction in the total area of the survey.

\subsection{Sky subtraction}
\label{subsec-sky}

 After the images have been masked, a large gradient in the sky level
persists along the direction of readout of the scans ($\alpha$). The
gradient is the result of temporal variation in the sky brightness
during the observations. This component of the sky is generally
smoothly varying; however, sharp changes in the sky level do
exist. These sharp changes can be due to bleed trails, scattered light
from bright stars in the field of view, or internal reflections in the
optics.  Both the smooth and sharp components can result in sky
brightness fluctuations of $\Delta\mu\sim$0.3 mag arcsec$^{-1}$.  In
the absence of sharp edges, an effective way of removing the sky is to
average the columns in a scan and then apply Savitsky-Golay
filtering. This approach minimizes random small-scale noise while
preserving the slowly varying component of the sky that we are trying
to model.  Unfortunately, if there are any sharp changes in the sky
level the Savitsky-Golay filtering also induces large errors in the
modelled sky level near the edges.  To circumvent this problem, a very
simple edge detection algorithm is employed prior to Savitsky-Golay
filtering.  Once edges have been detected and the magnitude of the
change has been determined, a step function representation of this
component of the sky level is constructed. This discontinuous
component is subtracted from the image, leaving only the smoothly
varying component of the sky, which can then be modelled with the
approach described above.  Figure \ref{fig:skyfit} shows sky fits for
a set of scans at the same right ascension, illustrating the procedure
described above. The presence of a discontinuous change in sky level
can be seen in the fourth fit from the top.  As a precaution, we also
mask all data within 20 pixels of detected discontinuities.
Subsequent to this fitting procedure, all images are normalized to
have the same median sky level.

Temporal effects are not the only source of large scale variation in
the sky level. In addition, two-dimensional structure is expected to
be present due to galactic cirrus, reflection nebulae, scattered
light, and saturated bright stars.  To correct for these large scale
variations we employ boxcar smoothing.  Each scan is smoothed on a
scale of 140$\arcsec$, and then the smooth, large-scale component of
the sky is subtracted.  This scale is sufficiently large that high
redshift clusters have a negligible contribution to the smoothing.

\subsection{Registration}

  An important step in this survey is registration of the scans.
Registration is required for several reasons, each of which requires a
different degree of accuracy. First, the cleaned images must be
aligned with an accuracy greater than the scale of the smoothing
kernel, or else cluster fluctuations will be diluted by the
misalignment. For our smoothing scale of 10$\arcsec$, the scans must
be aligned to $<$3$\arcsec$.  Second, we use the registration to
generate $\alpha,\delta$ coordinates for cluster candidates, which are
used for follow-up observations.  For this purpose, accuracy of a few
arcseconds is generally acceptable.  Finally, mosaics are constructed
from the images prior to object removal.
these mosaics should be aligned well enough that their profiles are
not bimodal although the cluster identification criteria are fairly
robust to misalignment. Typical seeing during observations was between
1 and 1.5$\arcsec$, so the alignment should be good to $<$1$\arcsec$.
Considering that each scan is 3.8$^{\circ}$ (20,000 pixels) long and
that the curvature of a scan varies with declination, subarcsecond
accuracy is a challenging requirement.

  To register the images, the IRAF routine DAOFIND is used to locate
all bright stars in a scan. Using an initial estimate for the scan
center and curvature of the scan (see Appendix A for a discussion of
projection effects for Great Circle scans and regular scans), these
stars are cross-correlated with the HST Guide Star
Catalog.{\footnote{The Guide Star Catalog was produced at the Space
Telescope Science Institute under U.S.  goverment grant. These data
are based on photographic data obtained using the Oschin Schmidt
Telescope on Palomar Mountain and the UK Schmidt Telescope.}}  Actual
values for the scan center and the curvature are determined by
maximizing the correlation with the guide stars. This
cross-correlation is done for each scan, so that the scans are
directly tied to a global coordinate system, which prevents the
accumulation of error inherent in using relative offsets. The
resulting alignment accuracy between scans is acceptable.  Near the
centers of scans, the alignment is generally good to
$<$0$\,\farcs$5. This quality degrades towards the ends as small
errors in curvature are amplified.  In the worst regions a maximum
error of 1$\farcs$5 is reached; however, generally the error at the
ends of scans is $\sim$ 1$\arcsec$.

\subsection{Mosaicing}

  The code we use to mosaic the scans is a modified version of the
DIMSUM package for IRAF.{\footnote{The DIMSUM package can be obtained
from ftp://iraf.noao.edu/contrib/}} Two mosaics are constructed at
each right ascension - one of the scans prior to object removal and
one after object removal. Our only noteworthy modification to the
DIMSUM package is with regard to the way in which masking is handled.
Masking was used for:
\begin{enumerate}
\item Bright stars and galaxies, for which cleaning is insufficient.
\item Bleed trails and sharp changes in the sky level, near which the
sky fit may be poor.
\item Image edges, where FOCAS may fail to detect some objects. 
\end{enumerate}

When mosaicing, masking can be handled in two ways: one can either
require that a region be unmasked in both overlapping scans for it to
remain unmasked, or one can require that it be unmasked in only one of
the scans and use the pixel values from the unmasked scan. The first
option is preferrable for objects intrinsic to the image (e.g. bright
stars), and also has the advantage of maintaining uniform survey
depth. However, the second option is useful if the region being masked
is small or narrow, in which case continuity may be more important
than uniform survey depth (particularly for FOCAS object detection).
We utilize the first approach for bright objects, bleed trails, and
sharp changes in sky level, but use the second approach for scan
edges.

\subsection{Smoothing and Detection} 
We next convolve the cleaned images with a smoothing kernel to
increase signal-to-noise prior to cluster detection.  Optimal
signal-to-noise is achieved when both the kernel profile and scale are
matched to cluster properties \citep[see][for a thorough discussion of
the topic]{phi91}. For an exact match, the surface brightness value in
the convolved image will correspond to the central surface brightness
of the object, while mismatch will lower the observed central surface
brightness.  We expect some degree of mismatch in this survey both
because we are probing a range in redshift and cluster mass (and hence
scale), and also because it is doubtful that high-redshift clusters
\begin{inlinefigure}
\epsscale{1}
\plotone{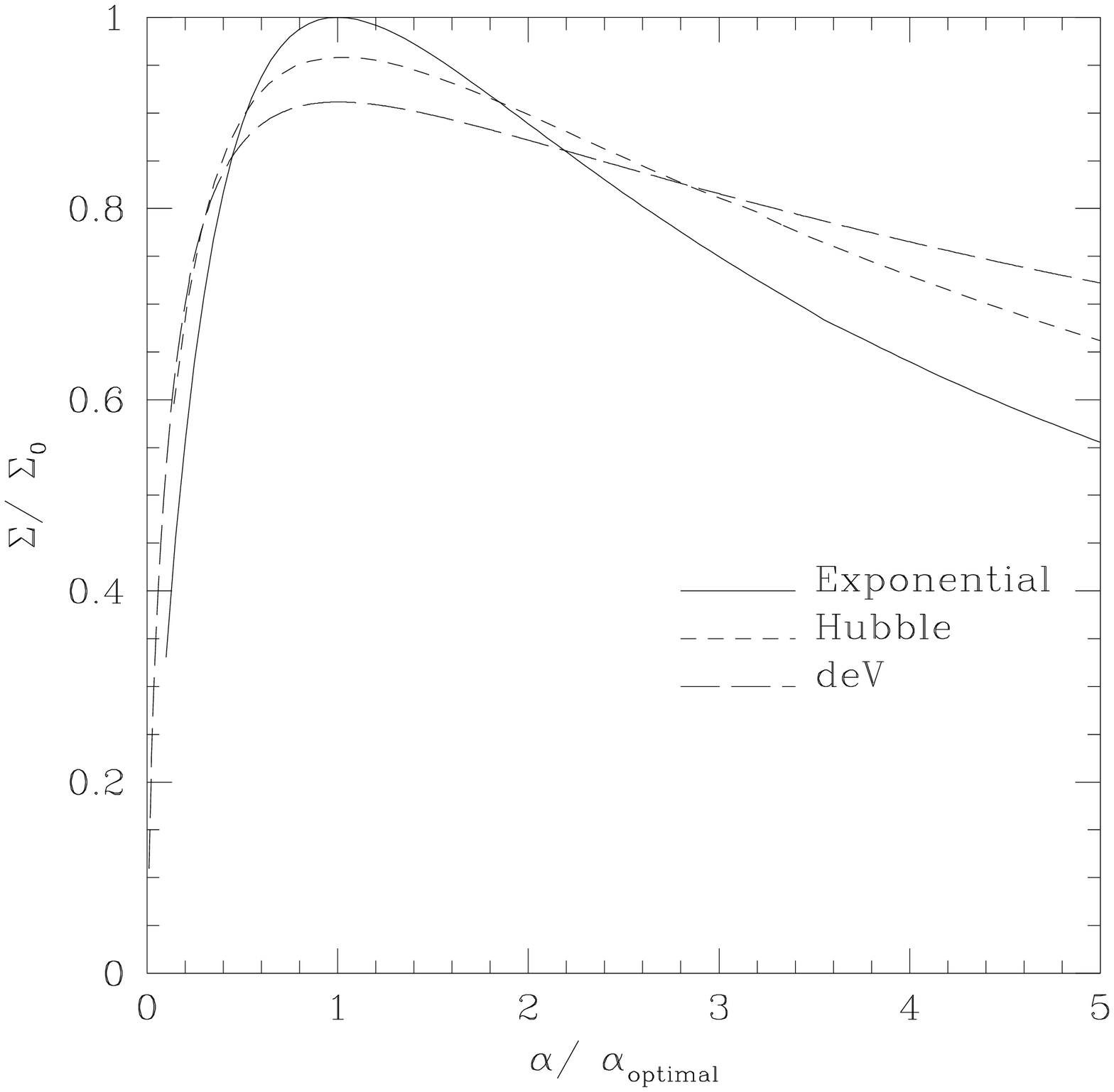}
\figcaption[filterimpact.eps]{Fractional decrease in
the observed surface brightness as a function of the relative mismatch
between the filter scale length, $\alpha$, and the optimal scale
length, $\alpha_{optimal}$. The curves shown correspond to an
exponential filter convolved with exponential, Hubble, and de
Vaucouleurs profiles.  \label{fig:filterimpact}}
\end{inlinefigure}

\noindent 
exhibit symmetric, uniform surface brightness profiles. Further, we
choose to employ an exponential kernel for the LCDCS - which is
probably not the correct profile for high-redshift
clusters.\footnote{Our use of an exponential kernel is motivated by
several factors. First, it is unclear what may constitute a "typical"
profile at high-redshift, where clusters may be far from virial
equilibrium. An exponential has the advantage of providing only
minimal degradation to do mismatch for a range of profiles including
Hubble and de Vaucouleurs, and also gives a bit less weight to large
radii (where the signal-to-noise is lower) than a Hubble
profile. Second, in absence of a single preferred profile, the
exponential is convenient since we also intend to use this data set to
construct a catalog of low surface brightness and dwarf galaxies.}
Consequently, it is useful to assess the impact of these factors upon
our sample.

To quantify the expected degradation due to our choice of kernel, we
compute the maximum cross-correlation for convolution of an
exponential kernel with Hubble and de Vaucouleurs profiles.  If the
scale of the kernel is optimally matched to the scale of the profile
(see below), then use of an exponential kernel results in net
decreases in the observed peak surface brightness of 4\% and 9\%,
respectively, for these profiles. Thus, we expect that our use of an
exponential kernel has minimal impact upon our detection efficiency.
Of greater importance is the choice of scale length for the kernel.
Figure \ref{fig:filterimpact} shows the fractional decrease in the
observed surface brightness for various profiles as a function of the
ratio of the actual kernel scale length to the optimal kernel scale
length. For all profiles, the fractional loss is $<$15\% when the
scale is matched to within a factor of two.

For the LCDCS, we set the scale length of our kernel to optimize our
ability to detect clusters at $z=1$.  If we assume a Hubble profile
for the cluster, then optimal scale length for our kernel is
$\alpha=0.6 r_c$, where $r_c$ is the Hubble core radius.  (This is
equivalent to setting the scale length such that both the exponential
and the Hubble profile fall off to half the central surface brightness
at the same radius). For rich clusters typical core radii are 
\begin{inlinefigure}
\plotone{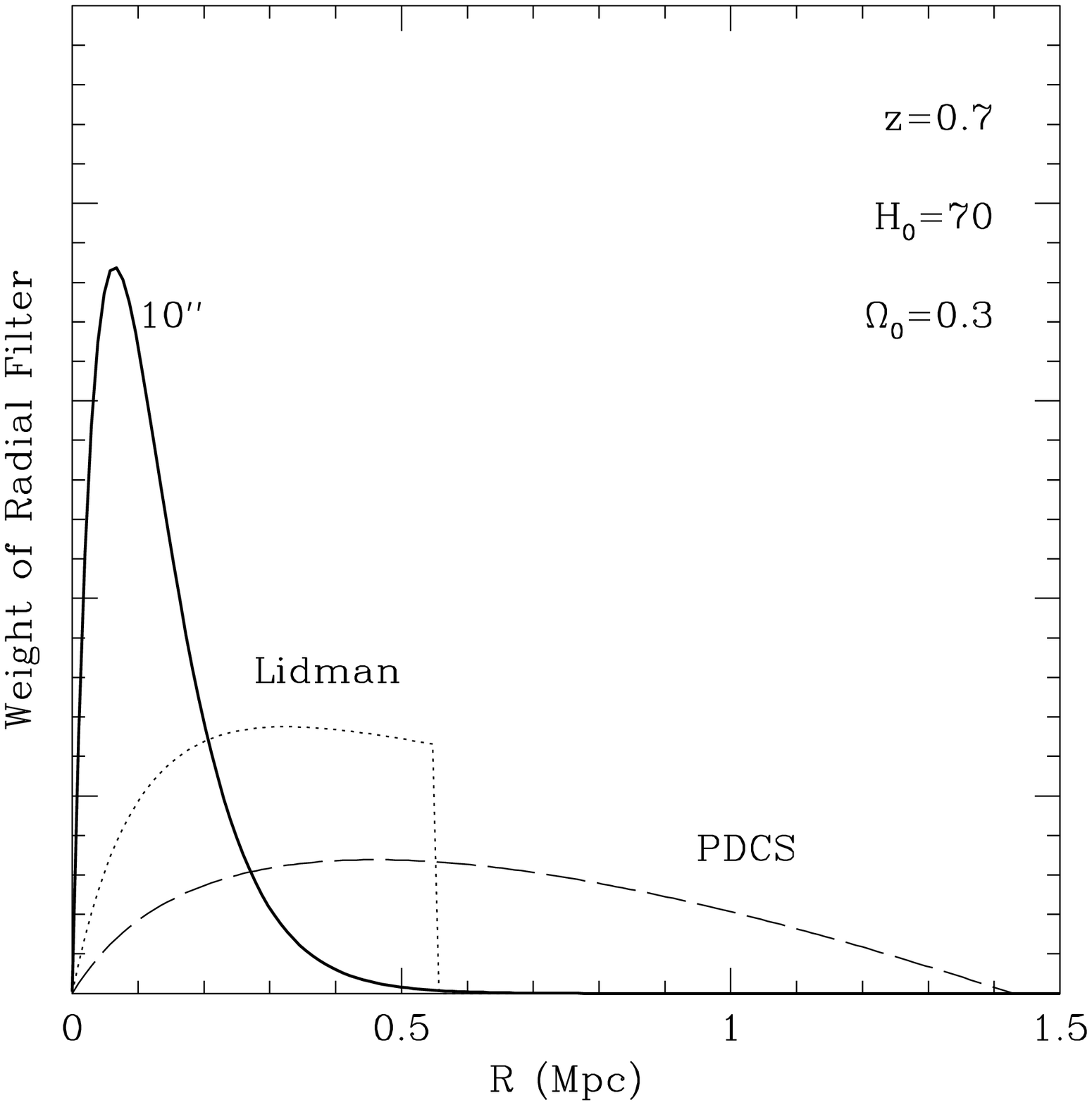}
\figcaption{ Radial filter weights from the
LCDCS, PDCS \citep{pos96}, and an optical survey by \citet{lid96}. The
filter weight is defined as $r f(r)/\int r f(r) dr$, where $f(r)$ is
the filter profile.  For the LCDCS most detection power is contained
within 300 kpc at $z$=0.7. In contrast, the other surveys shown, which
use galaxy counts, have most of their detection power at much larger
radii.  \label{fig:weight}}
\end{inlinefigure}

\noindent $r_c\sim
100-150 h^{-1}$ kpc, so the kernel size should be 60-90 $h^{-1}$ kpc
(12-18$\arcsec$ at $z=1$ for an $\Omega_0=0.3$ open model, or
11-16$\arcsec$ for an $\Omega_0=0.3$ flat model). Since the core radii
for clusters at this redshift may be smaller than local systems, we
choose to use a slightly smaller scale length of 10$\arcsec$.  This
choice of smoothing kernel does yield a factor of three mismatch at
$z=0.4$, which corresponds to about a 20\% fractional loss for Hubble
or de Vaucouleurs profiles.

The extent of the kernel used for this survey is significantly smaller
than those utilized for other optical surveys, with all the power
coming from within 250 $h^{-1}$ kpc (see Figure \ref{fig:weight}).
Because other surveys rely on number counts, by necessity their
kernels typically have scale lengths of order 1 Mpc in order to
include enough galaxies to generate a statistically reliable
signal. Unfortunately, use of such a large kernel increases the
likelihood of inclusion of poor systems due to projection effects or
large scale structure.  Our small kernel minimizes projection effects
and, because our method can detect cluster cores in the absence of a
well-formed envelope, we have a greater probability of detecting
clusters at early evolutionary stages.  Conversely, it is possible
that this technique will produce multiple, distinct detections for
clusters with significant substructure. While this effect is a
concern, we anticipate that it is a minor issue because we see no
enhancement in the angular correlation function at small
separations. Another possibile concern is whether we detect subclumps
of dwarf galaxies in nearby Abell clusters. Again, we see no evidence
of correlation between the locations of our candidates and Abell
clusters in the survey region (see Figure \ref{fig:skydist} in \S
\ref{sec-catalog}), and therefore expect that such detections do not
significantly impact the catalog.

Once a smoothed version of a scan region has been generated,
SExtractor v2.1.6 \citep{ber96} is employed to detect positive
brightness fluctuations.  SExtractor is used instead of FOCAS because
of its computational efficiency and a superior ability to deblend
overlapping detections. The detection threshold is set such that all
fluctuations with peak surface brightness in excess of $\mu_W$=28 mag
arcsec$^{-2}$ ($\sim5.4\times10^{-3}$ counts s$^{-1}$ arcsec$^{-2}$)
are identified.  For a typical scan set covering 7.6 square degrees
this procedure yields $\sim$2500 detections. Of these, roughly 65\%
are in the 5.9 square degree overlap region used for the cluster
survey (of which $\sim$30\% is typically masked).{\footnote{The
overlap region constitutes 78\% of the survey area, but only has 65\%
of the detections. This difference is a direct result of the lower
signal-to-noise ratio outside the overlap region leading to more
spurious detections.}}  About 5\% of these detections are galaxy
clusters at $z$$>$0.3; the rest are contamination. We use an automated
classification technique, described in the next section, to identify
viable cluster candidates.

\section{Cluster Identification}
\label{sec-id}
A critical element in creating the cluster catalog is identification
of viable cluster candidates and rejection of other sources of surface
brightness fluctuations.  Several techniques were explored for
isolating cluster candidates including hierarchical, divisive, and
fuzzy clustering algorithms \citep{gor81,jai88} and automated Bayesian
classification \citep{che88,goe89}, with each method using as input a
generalized set of statistics characterizing the detection image.  A
drawback of these algorithms though is that the physical meaning of
the resulting classes is not always clear, nor is it straightforward
to reproduce these classes for simulated data or new samples
(particularly for the fuzzy and Bayesian techniques).  Consequently,
we instead choose a more intuitive and easily reproducible approach
for extracting cluster candidates that yields a success rate
comparable to the other approaches.  We identify classes of objects
that induce fluctuations and separately isolate these classes
utilizing specific properties unique to the given type of object.

To do so, we employ information from both the smoothed and original images.
From the smoothed maps, we use:
\begin{enumerate}
\item{The observed peak brightness, $\Sigma_{obs}$, of the detected
surface brightness fluctuation.  It is defined as the maximum surface
brightness value in any pixel in the fluctuation, and the units of
$\Sigma_{obs}$ are counts s$^{-1}$ arcsec$^{-2}$. }
\item{The concentration, $C$, of the observed surface brightness
fluctuation.  We define the concentration in terms of the fraction of
pixels above a given threshold for two regions - an inner, circular
region of radius 15$\arcsec$ and an outer annulus extending from
15-30$\arcsec$. $C$ is one minus the ratio of the fraction above the
threshold in the outer annulus, $f_o$, to the fraction above the
threshold in the inner region, $f_i$:
\begin{equation}
C\equiv1-\frac{f_o}{f_i}.
\end{equation}
The threshold is set such that $\Sigma_{threshold}$=4.2$\times10^{-3}$
\csa = 2/3 $\Sigma_{lim}$, the limiting surface brightness of the
final catalog (see \S \ref{sec-catalog}).}
\item{The fraction of the detection region that is masked. We measure
the fraction of masked pixels within a $1\arcmin\times1\arcmin$ square
region centered on the surface brightness fluctuation. }
\end{enumerate}

From the original images we obtain photometric data about objects
located near the fluctuation, which enables us to discern whether the
detection was caused by a low surface brightness galaxy, bright star,
or nearby galaxy rather than by a cluster. Specifically, we measure:
\begin{enumerate} 
\item{The 5$\arcsec$ aperture magnitude, $m_{ap}$, of the 
brightest object (lowest m$_{ap}$) located within 15$\arcsec$ of the peak 
of the surface brightness fluctuation detected in the smoothed data.}

\item{The aperture magnitude, total magnitude, central surface
brightness, and full-width half maximum of the non-stellar object with
the brightest $m_{ap}$ located near the detected fluctuation. In
identifying this object we consider all sources located within
15$\arcsec$ of either the peak or the centroid of the detected
fluctuation.  An object is designated non-stellar if the stellarity
index $S<$0.97 or $m_{ap}$$>$20.  The latter constraint reflects a
preference at fainter magnitudes to be conservative and potentially
include a few stars rather than reject galaxies as stars. As will be
described in \S \ref{sec-properties}, this aperture magnitude is used
to estimate cluster redshifts, and it is preferrable to occasionally
underestimate a redshift due to stellar contamination rather than
overestimating the redshift due to identification of the brightest
cluster galaxy as a star. The total magnitude, $m_T$, is mag\_best in
SExtractor.}

\item{The aperture magnitude, total magnitude, central surface
brightness and full-width half maximum of the object with the lowest
$m_T$ located withing 15$\arcsec$ of either the peak or the centroid
of the detected fluctuation. }

\item{The number of galaxies, $n_{gal}$, brighter than $m_{ap}$=22 mag
within 30$\arcsec$ of the fluctuation.}
\end{enumerate}

The quantities listed above prove sufficient for classification of
most detected surface brightness fluctuations. Prior to
classification, we eliminate all detections that lie in regions where
E($B$-$V$)$>$0.10 \citep{schl98}. This criteria removes all detections
at $\alpha$$>$15h12m, where the data are at low galactic latitude, and
an additional 32 detections at lower right ascension that would
otherwise be included in the statistical catalog (see
\S\ref{sec-catalog}).  We also eliminate all detections lying in
regions that, based on visual inspections of IRAS maps obtained from
SkyView \citep{mcg96} and the smoothed optical maps generated in this
work, contain galactic cirrus with significant structure. We do so
because galactic cirrus both reflects optical light and produces
extended red emission from 6000-8000\AA \citep{guh89,szo98}, and
nonuniformity in this emission/scattered light generates surface
brightness fluctuations that are indistinguishable from
cluster-induced fluctuations. An example of this phenomenon is shown
in Figure \ref{fig:iras}.  Finally, to preserve a constant threshold
for the survey, a target is only considered a cluster candidate if the
detection lies in a region that was observed twice during the course
of the survey. All other detections are flagged and removed from
further consideration.  We then proceed by identifying the physical
origins of the remaining fluctuations and sequentially isolating
classes of objects. The types of objects, and parameters used to
isolate them, are described below.

{\it{Stars}} - As described in $\S$ \ref{sec-redux}, all detected
objects in the drift-scan data are replaced with local, random sky
pixels. This procedure effectively removes most resolved objects;
however, for bright stars the `sky' pixels used to replace the star
often contain a non-negligible amount of light from the wings of the
stellar point spread function. As a result, a surface brightness
fluctuation is produced coincident with the star for stars with
14.5$<$m$_{ap}$$<$18.5. For m$_{ap}$$<$14.5 the stars are masked; for
m$_{ap}$$\ga$18.5 the stars are sufficiently faint that they do not
induce detectable fluctuations.  Fluctuations induced by bright stars
are cleanly isolated using the stellarity index and magnitude of the
brightest object near the fluctuation. We classify a detection as
being induced or strongly influenced by a star if the brightest object
within 15$\arcsec$ of the peak of the detection has a stellarity index
$S$$>$0.9 and $m_{ap}$$<$18.5.  Stars are responsible for half of all
detections.

{\it{Galaxies}} - Individual galaxies can also cause detectable
surface brightness fluctuations if they have significant flux at large
radii. As with bright stars, the local `sky' pixels used to replace
the core of the galaxy produce a surface brightness feature coincident
with the original location of the galaxy. To minimize contamination of
the cluster catalog by these detections, we impose two
criteria. First, we classify a detection as a galaxy-induced if there
is a galaxy with $m_T$$<$18.75 within 15$\arcsec$ of the peak or
centroid of the detection (which is about a quarter of a magnitude
brighter than a typical BCG at $z$=0.30).

We augment this procedure with a second cut that takes into account
the surface brightness of the detection and the number of galaxies
within 30$\arcsec$ of the detection. If the brightest galaxy near the
fluctuation is the brightest cluster galaxy (BCG), then the
fluctuation should be significantly brighter than one induced by an
isolated, local galaxy of the same apparent magnitude. In addition, at
$z$$\sim$0.3 we should be able to identify projected galaxy
overdensities associated with real clusters.  To discriminate
low-redshift clusters from isolated galaxies, we define the parameter
$p_{gal}$ such that
\begin{equation}
 p_{gal}=\sqrt{ (2.5*(\Sigma_{obs}/\Sigma_{det}-1))^2 + (n_{gal}-1)^2 },
\end{equation}
where $\Sigma_{det}$=5.4$\times10^{-3}$ \csa is the detection threshold of
the survey. Using this parameter, we classify detected fluctuations as 
galaxies if
\begin{equation}
p_{gal}<2.5-6(m_{ap}-20).
\end{equation}
This criteria is empirical, constructed from visual inspection and
follow-up imaging of candidates at 10h$<$$\alpha$$<$10h15m ($\sim$5\%
of the survey).  No detections are eliminated when $m_{ap}$$\ga$20.4,
which means that this cut has 
\begin{inlinefigure}
\plottwo{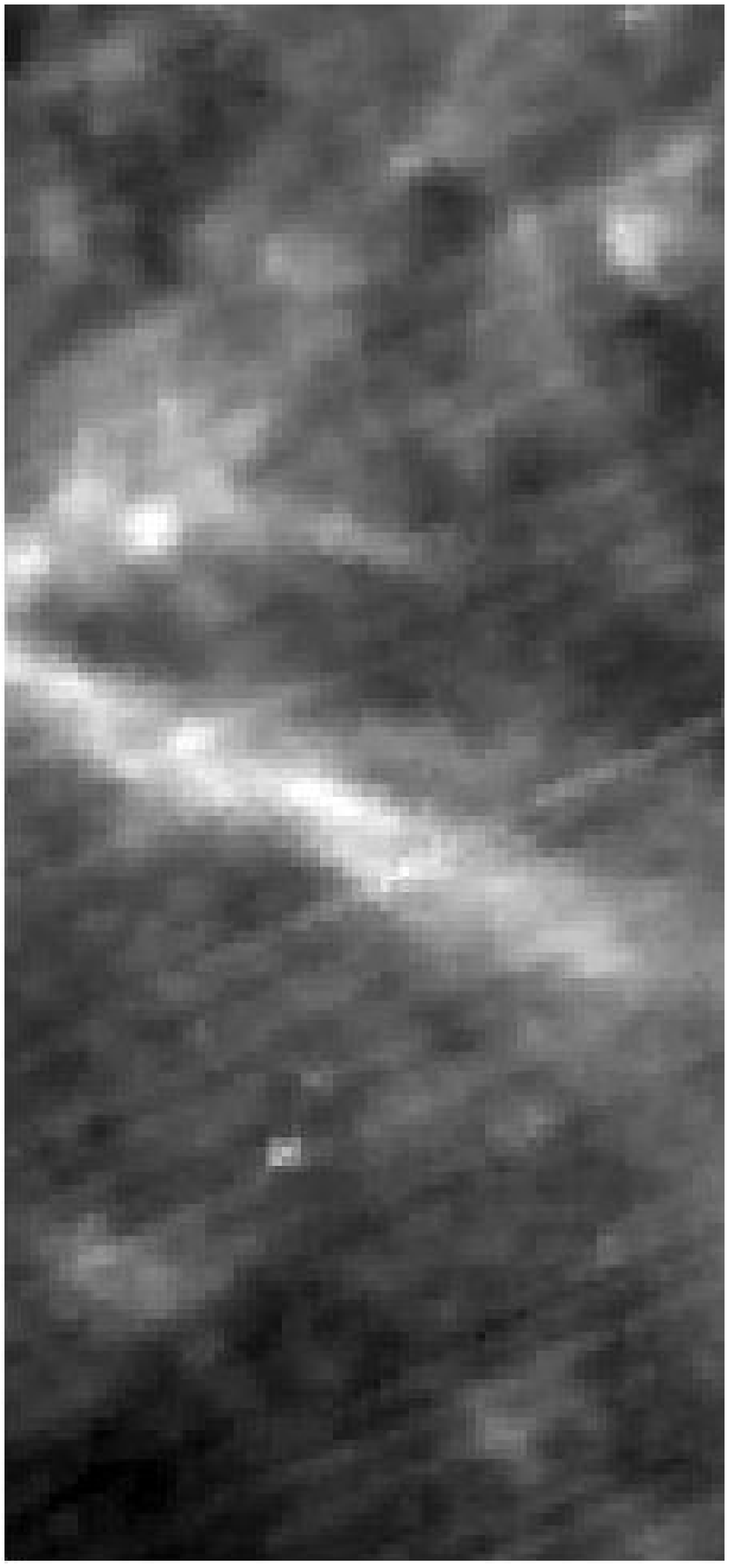}{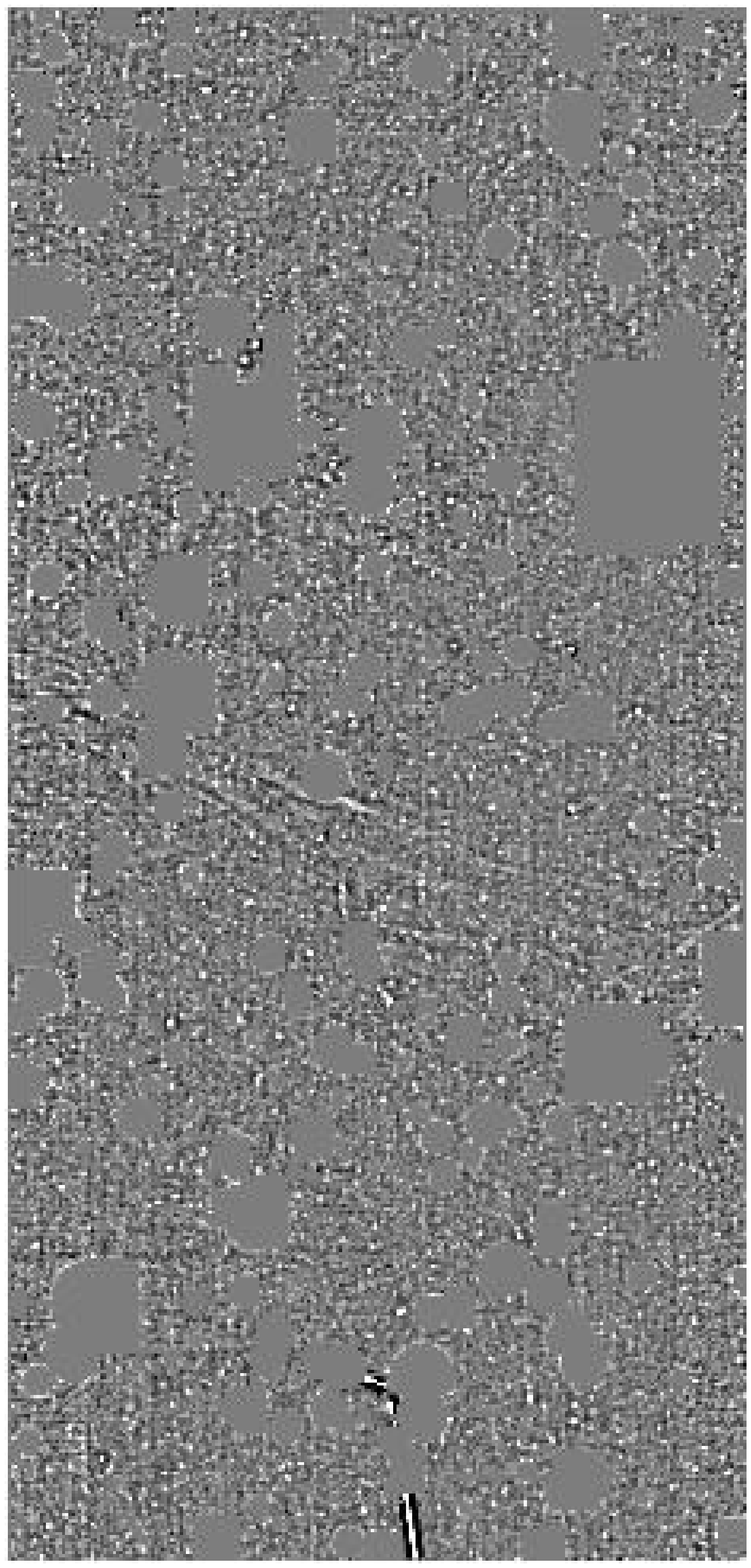} \figcaption[iras.eps]{ Comparison of an
IRAS 60$\mu$ image (left) and smoothed optical map (right) of a small
section of the survey. The images are 1.9$\degr\times$3.9$\degr$ in
size. East is up and north is to the right. Note the large cirrus
spike evident in the center of the IRAS image. In the optical image
the large scale structure of the spike is filtered during sky removal;
however, small scale features persist and can be seen in the optical
map. Consequently, via inspection of the IRAS and optical maps we
catalog these regions and reject all detections in these areas. Note
also the satellite trail at the lower edge of the optical
image. Satellite detections that survive the automated selection
criteria discussed in \S \ref{sec-id} are manually removed from the
final sample.\label{fig:iras} }
\end{inlinefigure}

\noindent no impact on the final catalog at
$z_{est}\ga$0.35 (see \S \ref{subsec-redshifts}).  Together the galaxy
identification criteria eliminate 14\% of all detections.

{\it{Low Surface Brightness and Dwarf Galaxies}} - After elimination
of fluctuations caused by stars and galaxies, we identify detections
induced by low surface brightness galaxies (LSBs) and dwarf galaxies.
To identify these galaxies, we rely on the SExtractor information from
the original images for the two galaxies within 15$\arcsec$ that have
the brightest aperture and total magnitudes. Qualitatively, if a
galaxy is a nearby LSB or dwarf, we expect that the central surface
brightness will by definition be low, the FWHM will be relatively
large since the galaxy is nearby, and the total magnitude will be much
larger than the aperture magnitude (due to the previous two traits).
We employ a two-stage approach to eliminating LSBs. First, we identify
detections as potentially induced by LSBs if $m_{ap}$-$m_T$$>$1.25 for
either of these two nearby, brightest galaxies. Second, we define the
parameter, $p_{LSB}$,
\begin{equation}
p_{LSB}=FWHM - 1.2 m_T +11.5,
\end{equation}
and eliminate detections as likely LSBs if $\mu_0$$>$22 mag
arcsec$^{-1}$ and $p_{LSB}$$>$0 for either of these two galaxies (see
Figure \ref{fig:lsbcut}).  This constraint, which is again empirical,
is calibrated via visual inspection.  A total of $\sim$250 detections
in the entire survey are classified as LSB candidates based upon these
criteria.  By visual inspection of these detections, we estimate that
$\sim$5 clusters with $\Sigma_{obs}$$>$$6.25\times10^{-3}$ \csa are
lost due to misidentification as LSBs. Conversely, $\sim$ 20 LSBs are
mistakenly identified 
\begin{inlinefigure}
\plotone{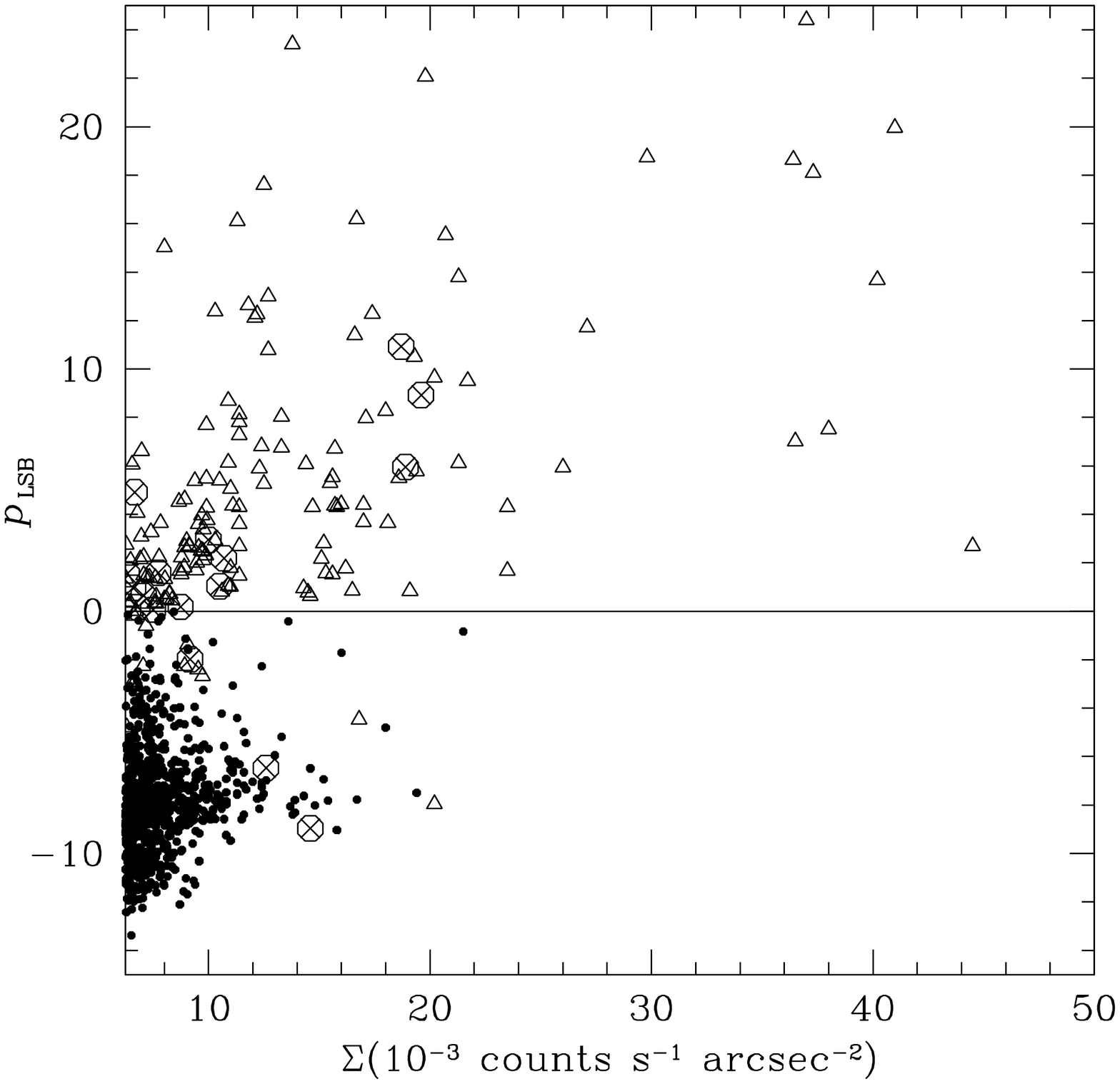} \figcaption[f10.eps]{ Criteria used to isolate
LSBs. Filled circles are clusters, triangles are low surface
brightness galaxies, and circled crosses are systems that are visually
ambiguous when classified by eye from the scans.  Detections with
$p_{LSB}$$>$0 are identified as probable LSBs. The data plotted
represent data from the entire survey. Only the first 10\% of the
survey region was used to calibrate this division, but all sources
designated as LSBs or clusters were visually inspected after automated
classification to verify the effectiveness of this
procedure. \label{fig:lsbcut}}
\end{inlinefigure}

\noindent as clusters and included in the final catalog.

{\it{Partially masked detections}} - We next eliminate from the
statistical sample all other detections for which greater than 10\% of
the detection region is masked. Such obscuration of the detection
region leads to difficulty in determining the true cause of a
detection, and consequently results in increased contamination of the
cluster sample.

{\it{Scattered light and assorted spurious reduction-induced signals}}
- Subsequent to the removal of all of the sources listed above,
$\sim$8\% of the total detections remain. Of these, the principal
remaining contaminants are sources arising from scattered light and
signals related to the reduction process. As a class, these detections
are characterized by low peak surface brightness and large spatial
extent relative to the clusters. To isolate these, we use the
concentration measure, $C$, as a discriminant.  Detections of this
type are on average less concentrated (lower $C$) than clusters of
comparable surface brightness. The division between the two classes is
shown in Figure \ref{fig:lsscut}.

{\it{Satellites and Bleed Trails}} - Satellites and bleed trails from
saturated stars, because they are not detected by FOCAS and replaced
with sky pixels, also induce fluctuations.  These sources are rare and
easily identifiable, and so they are identified by visual
inspection. This is the only aspect of the classification procedure
that is not automated.

{\it{Tidal Features}} - Another source of detections is tidal features
in nearby, interacting systems. We do detect fluctuations induced by
large tidal features; however, we do not identify these as a distinct
class of objects when generating the cluster catalog. Tidal features,
because of their varied morphologies, are difficulty to identify via
\begin{inlinefigure}
\plotone{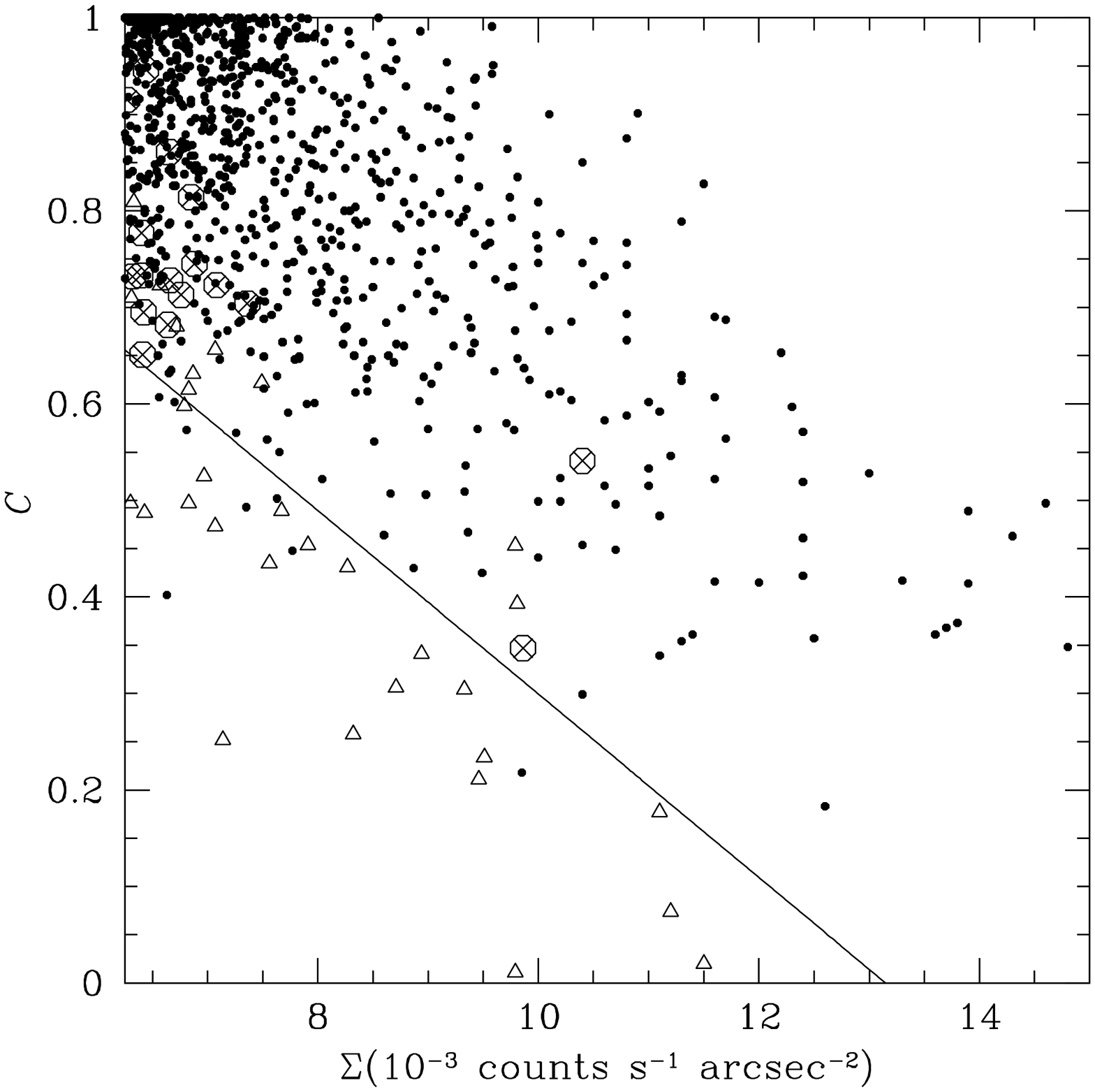} \figcaption[f11.eps]{ Criteria used to eliminate
spurious detections.  Filled circles are clusters, triangles are
spurious detections, and circled crosses are systems that are visually
ambiguous. The data plotted are drawn from the entire survey region.
\label{fig:lsscut}}
\end{inlinefigure}

\noindent automated criteria. Fortunately, however, the surface density of
interacting systems capable of inducing detections is low. From visual
inspection, we estimate the fractional contamination by tidal tails is
$\sim$1\% in the final catalog.

{\it{Chance Galaxy Superpositions}} - Chance superpositions of several
field galaxies with magnitudes near the survey limit provide a final
source of contamination. Such detections are indistinguishable from
true cluster detections, and so can only be eliminated with deeper
follow-up imaging.  While it is difficult to robustly estimate the
magnitude of this contamination directly from the survey data, these
superpositions may be the primary source of false detections in the
final catalog (see \ref{subsec-contamination} for observational
constraints on the net contamination rate).

{\it{Clusters}} - All of the remaining 2670 detections are cluster
candidates. As will be discussed in \S \ref{sec-catalog}, not all of
these systems are included in the final statistical catalog; however,
we consider all these detections to be viable candidates. The
properties of the sample are discussed in further detail in \S
\ref{sec-properties}.

\section{The Catalog}
\label{sec-catalog}

Having established our selection criteria, we now define the main,
statistical catalog for the Las Campanas Distant Cluster Survey.  This
statistical sample is designed to be an automated, reproducible
catalog with which to study properties of the cluster population as a
whole. For this catalog, we impose two additional constraints beyond
those discussed in \S \ref{sec-id}. First, we restrict the catalog to
detections with $\Sigma_{obs}$$>$$6.25\times10^{-3}$ \csa. This
restriction is due to relatively high contamination for fainter
systems, as determined by follow-up imaging. Second, we restrict the
sample to clusters at $z$$>$0.30. Below this redshift incompleteness
and confusion with individual galaxies become significant issues.

\begin{inlinefigure}
\plotone{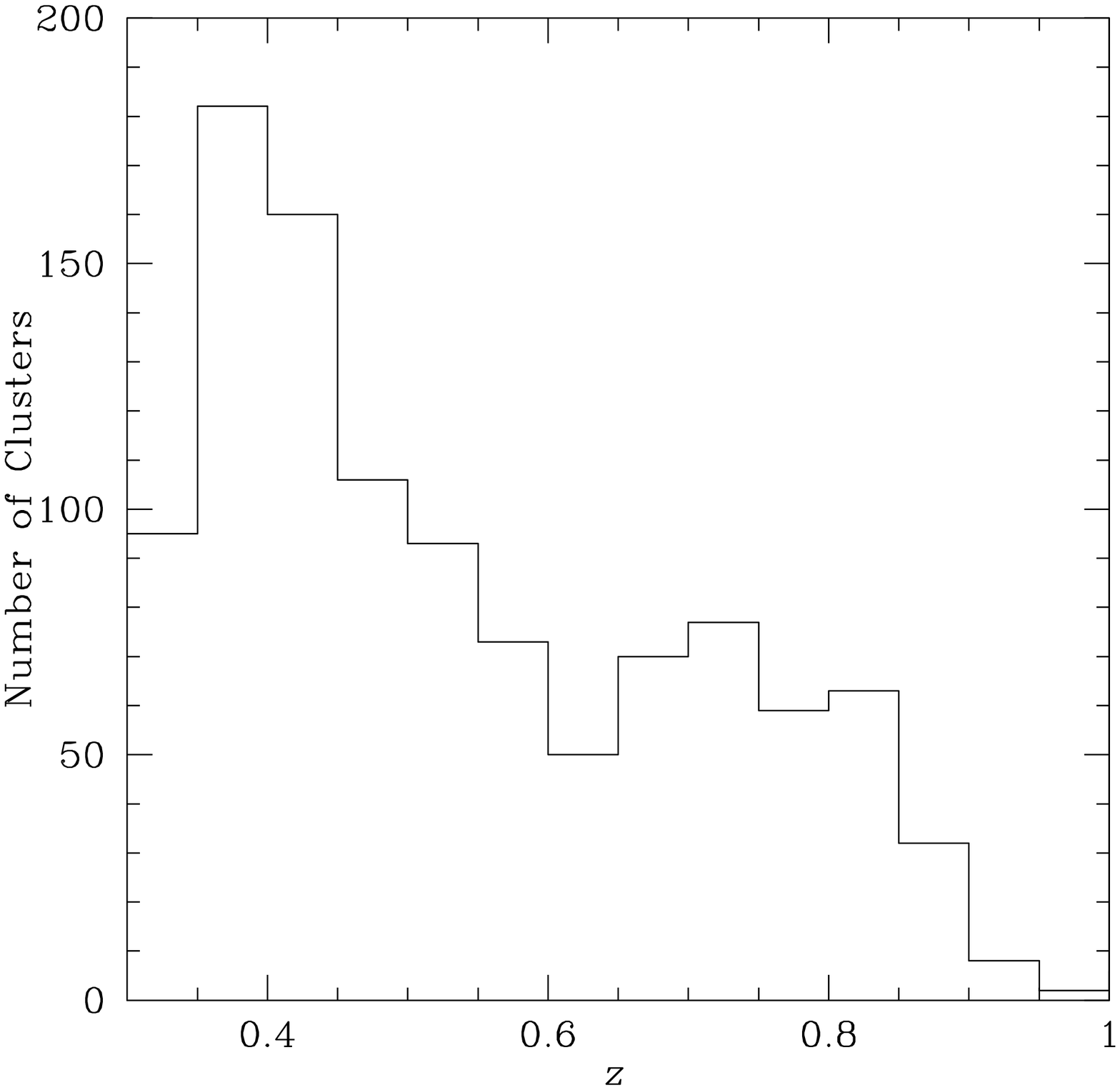} \figcaption[f12.eps]{ Redshift distribution of
clusters in the statistical catalog.
\label{fig:zdist}}
\end{inlinefigure}

There are a total of 1073 cluster candidates in the statistical
catalog.  These candidates are drawn from a net survey area, after
masking, removal of regions of high extinction, and rejection of data
in close proximity to bright stars and galaxies, of 69 square
degrees. Equivalently, the projected density of cluster candidates is
15.5 per square degree, which is comparable to the PDCS and EIS
catalogs (13.4 per square degree for the primary PDCS sample, and 21.1
per square degree for the full EIS sample).  As the redshift baselines
of these three surveys is slightly different, a more fair comparison
is the projected density of clusters at $0.3\le$$ z$$ \le$ 0.8. Within
this range, the LCDCS, PDCS and EIS find projected densities of 14.2,
10.8, and 10.7 candidates per square degree, respectively. The higher
projected density of the LCDCS is likely an indication that we probe
to slightly lower mass, but may also be partly due to higher
contamination (see \S \ref{subsec-contamination}).

\begin{figure*}
\epsscale{1.00}
\plotone{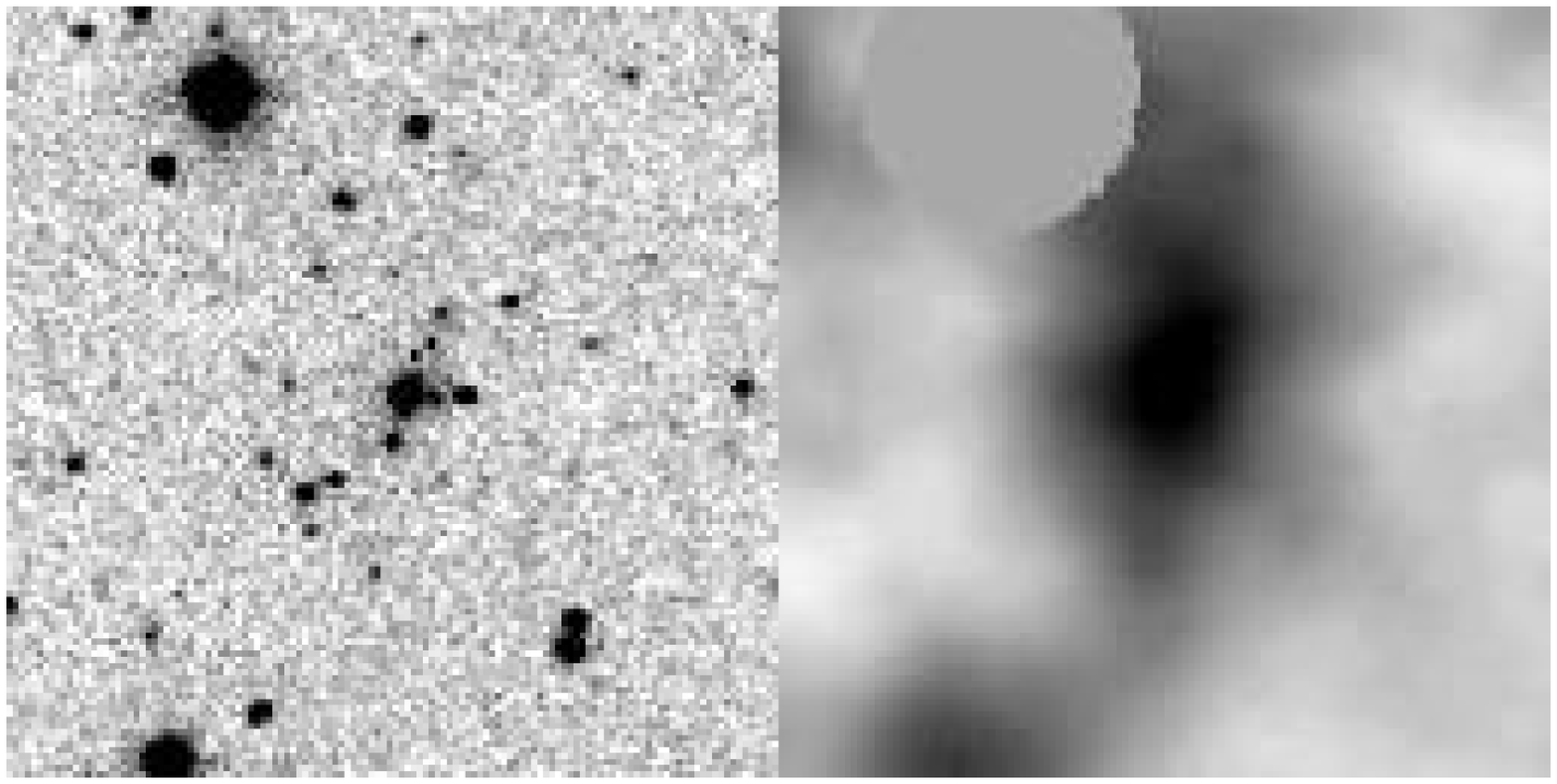}
\plotone{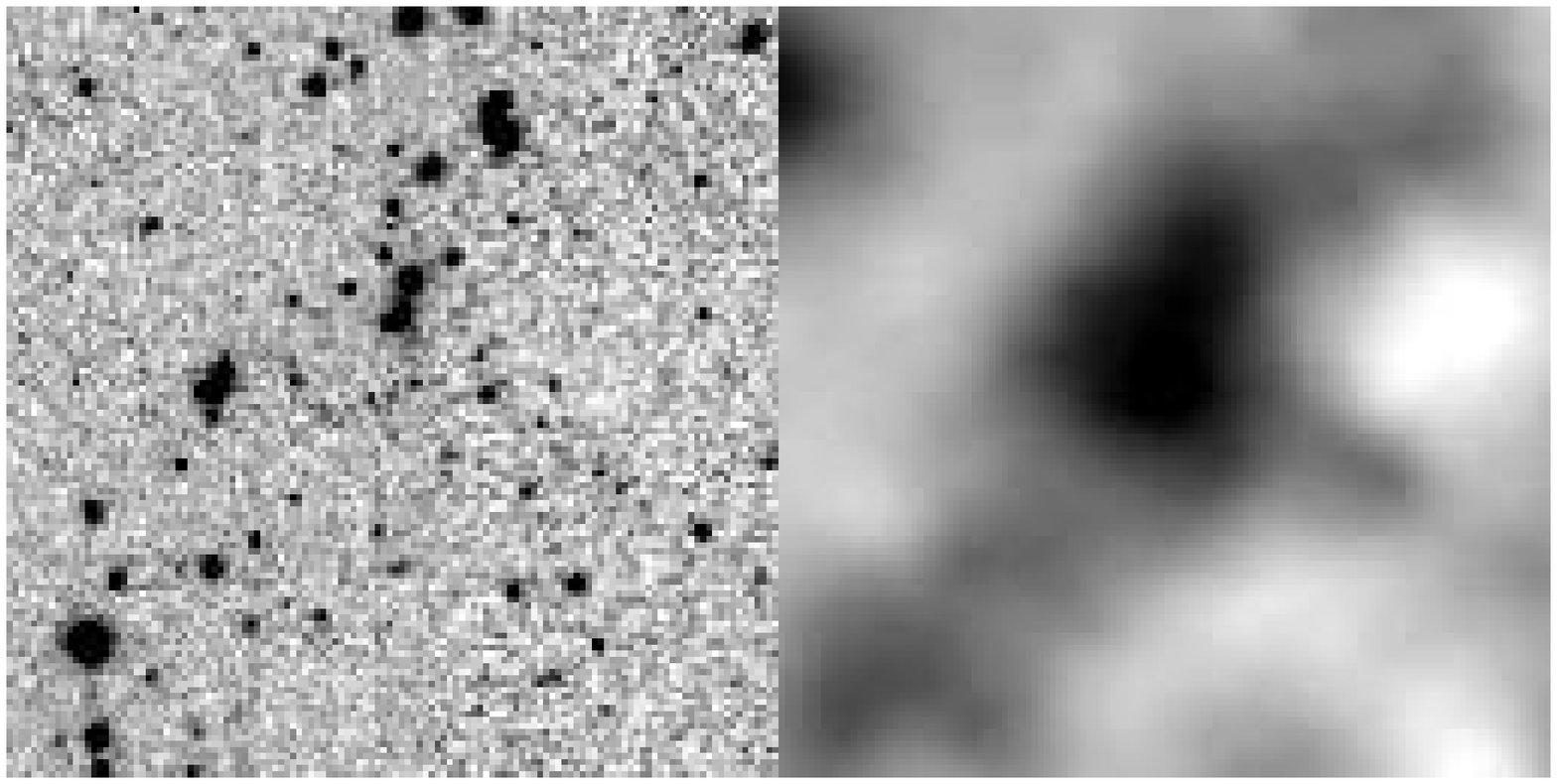}\\ 
LCDCS 0347 ($z_{est}$=0.30) \hskip 2.75in LCDCS 0169 ($z_{est}$=0.39)
\vskip 0.2in
\plotone{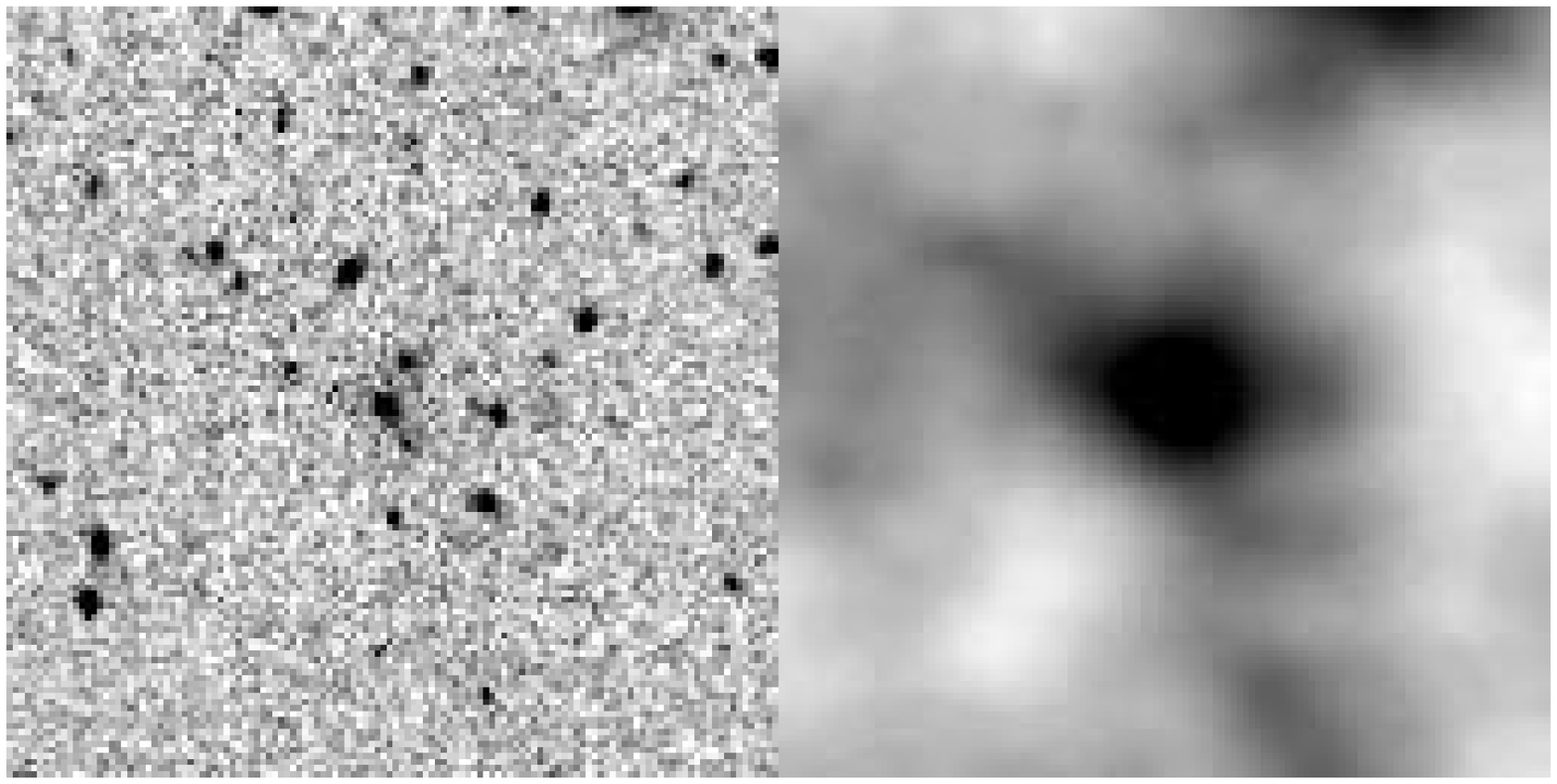}
\plotone{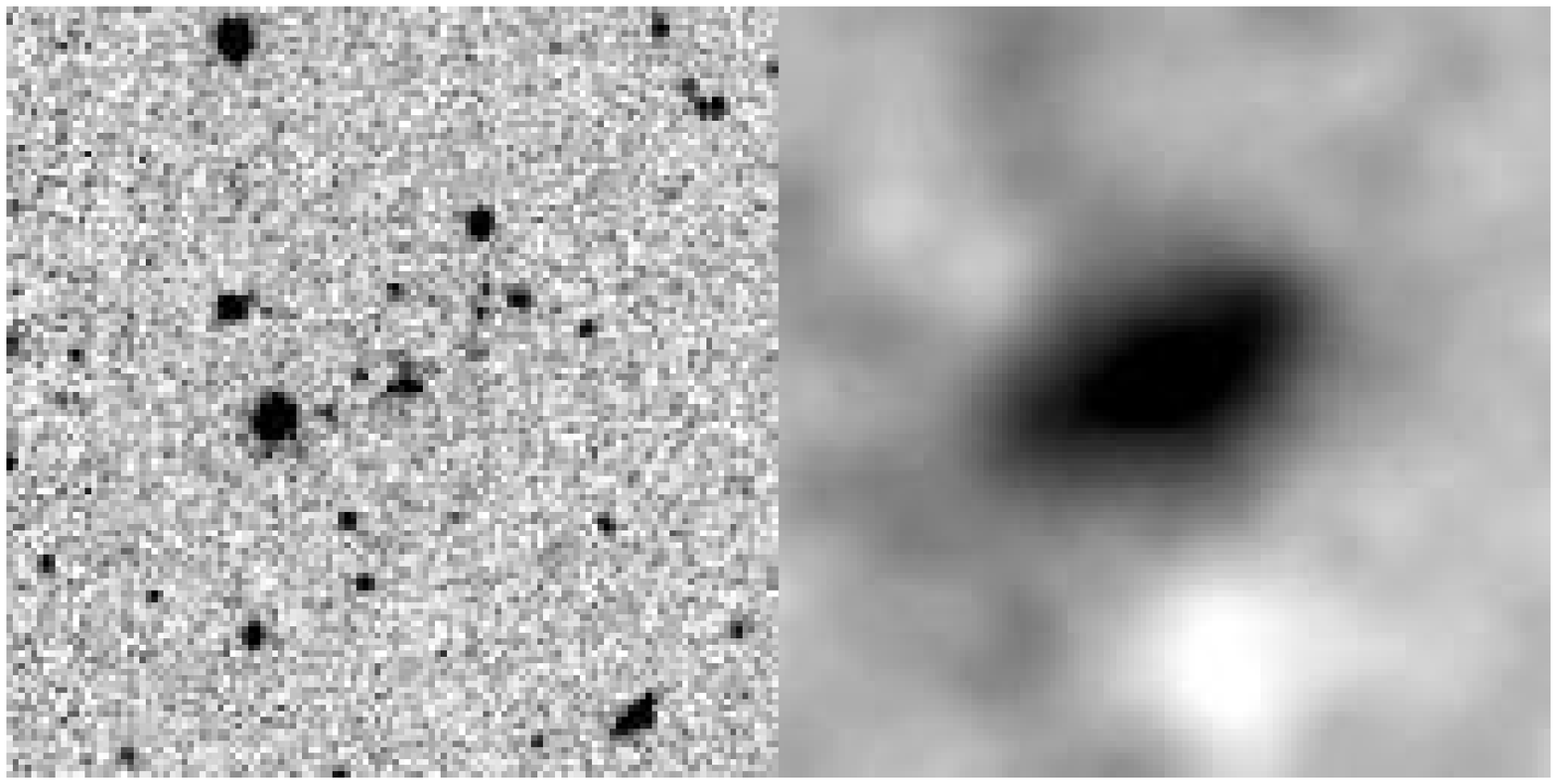} \\ 
LCDCS 0589 ($z_{est}$=0.50) \hskip 2.75in LCDCS 0795 ($z_{est}$=0.59)
\vskip 0.2in
\plotone{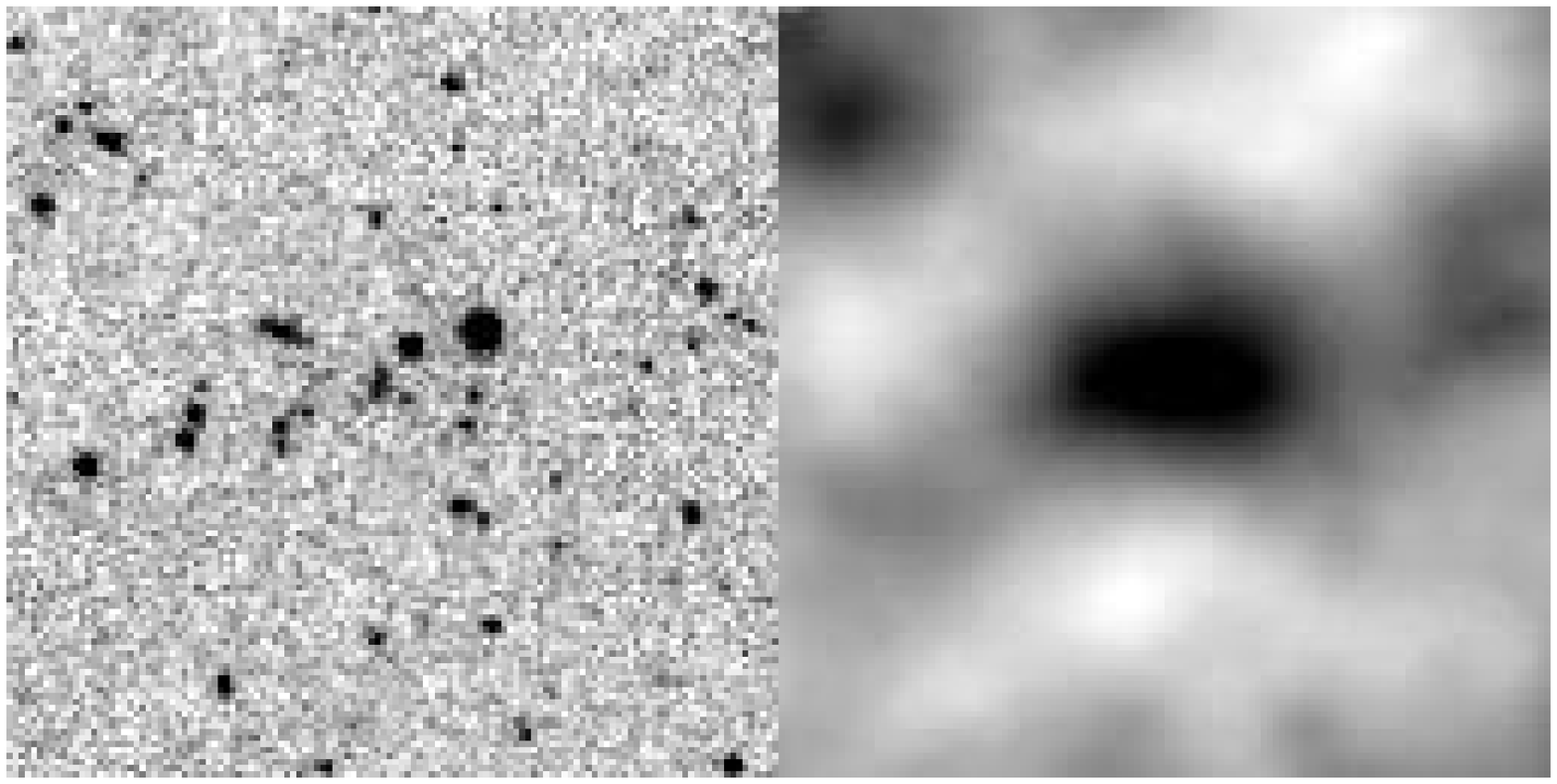}
\plotone{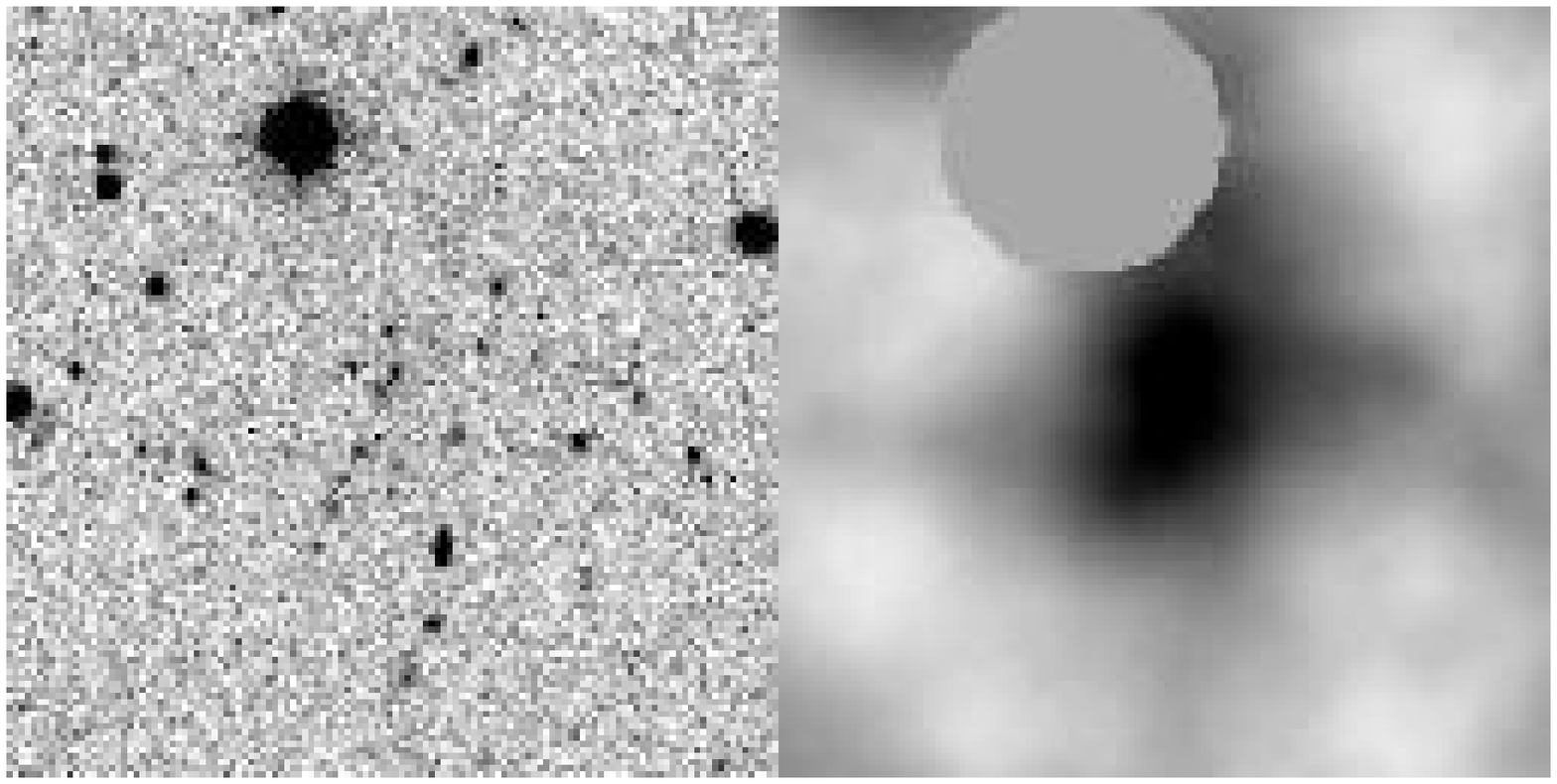} \\ 
LCDCS 0827 ($z_{est}$=0.71) \hskip 2.75in LCDCS 0130 ($z_{est}$=0.80)
\vskip 0.2in 
\plotone{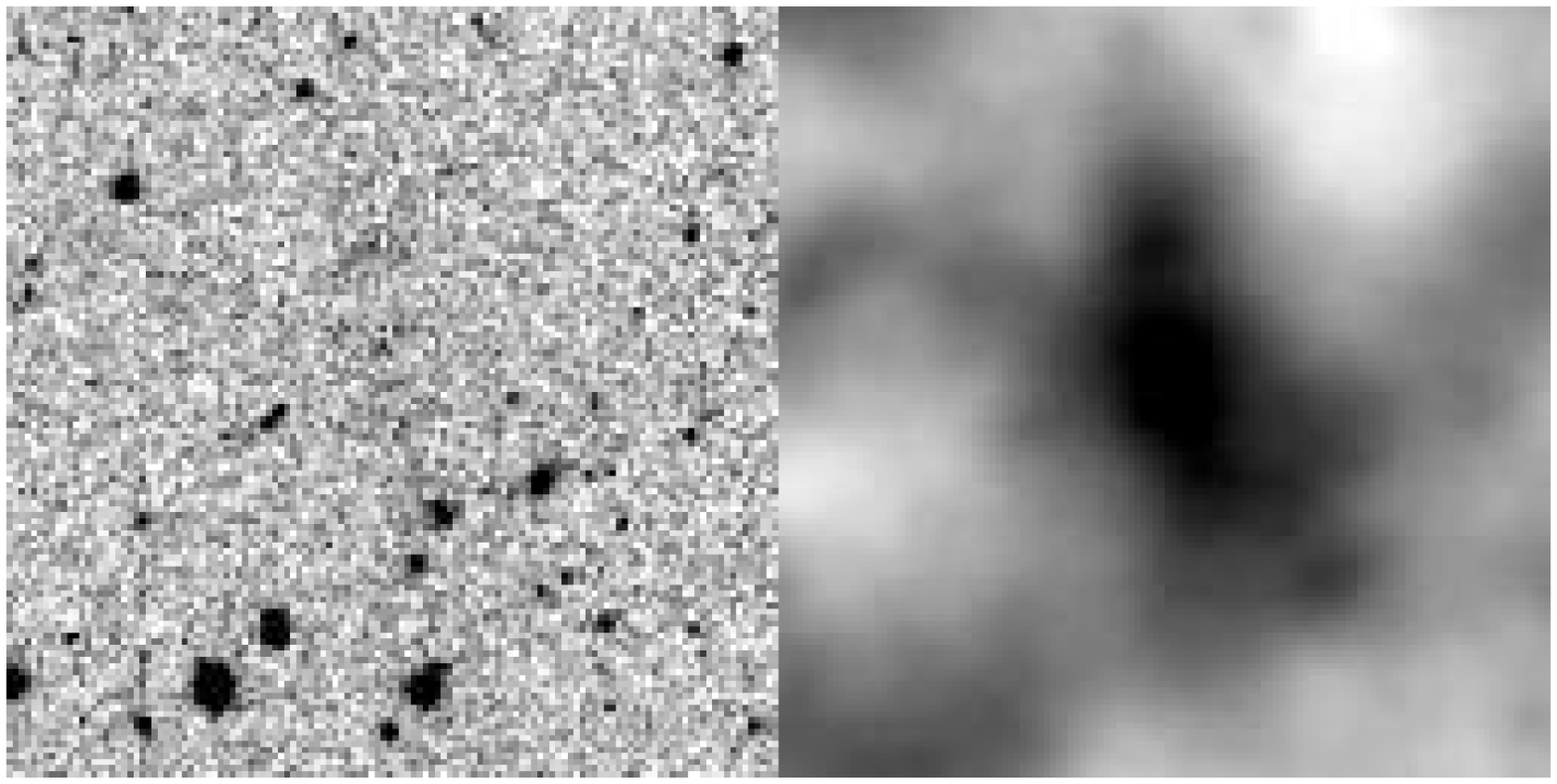} \\ 
LCDCS 0797 ($z_{est}$$>$0.85)
\epsscale{1} 
\figcaption[lcdcs130.eps]{ Detection images for seven of
the candidates in the LCDCS catalog. The candidates shown were chosen
to uniformly span the redshift range of the survey. Further, all
candidates shown have $\Sigma_{obs}$=9-9.5$\times$10$^{-3}$ cts
s$^{-1}$ arcsec$^{-2}$, except for LCDCS 0797
($\Sigma_{obs}$=7.9$\times$10$^{-3}$ cts s$^{-1}$ arcsec$^{-2}$).
\label{fig:clusterimages}}
\end{figure*}

Table \ref{table:statistical}
lists all candidates in the catalog. For
each candidate, we provide an identification number in Column 1, right
ascension in Column 2, and declination in Column 3. Estimated
redshift, the derivation of which is described in \S
\ref{subsec-redshifts}, is given in Column 4. Subsequent columns list
observed surface brightness, $\Sigma_{obs}$, in 10$^{-3}$ counts
s$^{-1}$ arcsec$^{-2}$ (Column 5), extinction corrected surface
brightness, $\Sigma_{cor}$, in 10$^{-3}$ counts s$^{-1}$ arcsec$^{-2}$
(Column 6), and the $\bv$ extinction from \citet{schl98} (Column
8). For the extinction correction, we take $A_W$=3 $E$($\bv$), which
is the average of the published values for $A_V$ and $A_R$ (see Figure
\ref{fig:filter} for a comparison of $W$ with these passbands).
Finally, Column 9 contains additional information for some candidates,
including notes for objects that from inspection are obviously not
clusters. All observed surface brightnesses have an associated
uncertainty of 1.6$\times$10$^{-3}$ \csa (see \S
\ref{subsec-completeness}); derivation of the uncertainty for the
extinction-corrected values is straightforward.  Figure
\ref{fig:zdist} gives the redshift distribution of these candidates,
and Figure \ref{fig:skydist} shows the projected distribution overlaid
with the locations of Abell clusters.  In Figure
\ref{fig:clusterimages} we show detection images for seven candidates
covering the redshift range of the survey.

The statistical catalog is the primary sample presented in this
paper. However, because the interests of researchers who use this
catalog may be diverse, we have also constructed a smaller,
supplemental catalog that contains additional cluster
candidates. These supplemental targets are detections that failed one
or more of the automated criteria for inclusion in the statistical
sample, but visually appear to be highly probable cluster
candidates. For this supplemental catalog, we maintain the restriction
$\Sigma_{obs}$$>$6.25$\times10^{-3}$ \csa, and no attempt was made to
recover clusters that were mis-classified as stars, or detection for
which more than 50\% of the detection region was masked. We do,
however, re-evaluate all detections rejected on the basis of
extinction, lesser masking, identification as an LSB, or
identification as a spurious feature. We also relax the galaxy
rejection criteria such that clusters may have $z$$<$0.30 and
$m_T$$<$18.75. From this extended sample we visually identify 112
additional cluster candidates. Table \ref{table:supplemental} 
lists
the supplemental targets, including the reason for rejection from the
statistical sample. We note that the $\Sigma_{obs}$ values of clusters
rejected due to masking should be used with caution, as partial
masking can skew this quantity. The redshift distribution of the
supplemental catalog is comparable to the statistical catalog for
$z$$\la$0.65. At higher redshift there are more candidates in the
statistical catalog, due both to contamination in the statistical
catalog and failure to visually identify high redshift candidates for
the supplemental catalog.

\section{Sample Properties}
\label{sec-properties}

To maximize the utility of this sample, we must constrain four key
properties of the catalog. Specifically, we need to determine the
completeness of the sample, the contamination rate, the redshift
distribution, and the mass limit of the survey as a function of
redshift. Further, if possible, we wish to estimate redshifts and
masses for individual systems directly from the survey data. We
address each of these issues below.

\subsection{Completeness}
\label{subsec-completeness}

 To quantify the completeness of the sample and uncertainties in
measured parameters, we run simulations in which a dozen of the
detected cluster candidates - including several below the surface
brightness threshold of the statistical catalog - are randomly
reinserted into the survey data.  Information about these clusters,
which were selected to span a range of surface brightness and
redshift, is given in Table \ref{table:completeness}.  
These simulations are specifically aimed at measuring completeness as a
function of surface brightness and scatter in the observed values of
$\Sigma_{obs}$ for clusters similar to the ones that we detect in the
LCDCS.  A corollary concern is whether we systematically miss any
subset of the cluster population. If so, then the completness rate
derived from these simulations will overestimate the actual value.
Figure \ref{fig:filterimpact} implies that our ability to detect
clusters is not strongly dependent upon the core radius. In addition,
we have obtained survey quality drift-scan imaging of 17 clusters at
$z=0.35-0.85$ drawn from published optical and X-ray catalogs to test
for any 
\begin{inlinefigure}
\plotone{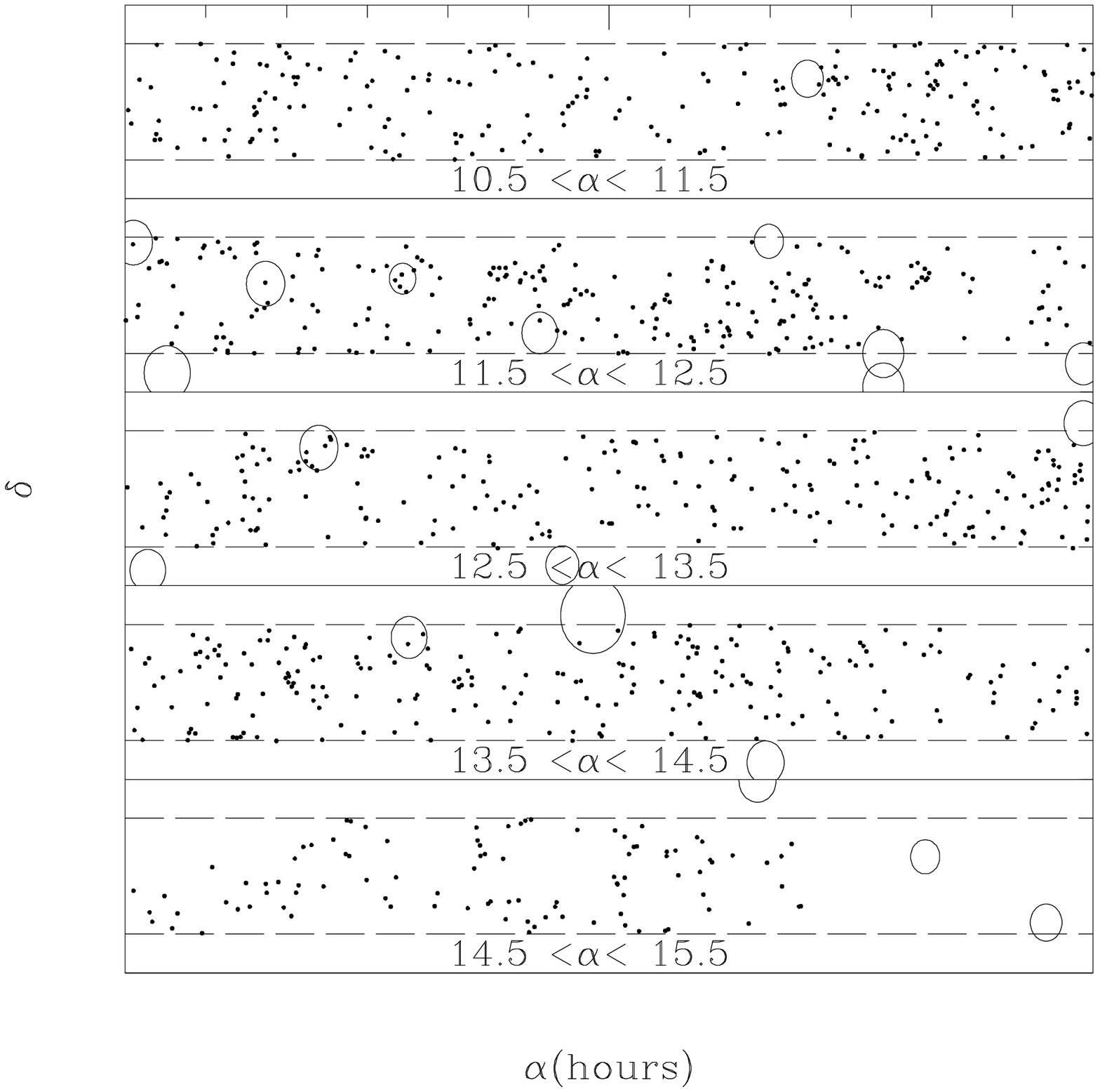} \figcaption[skydist.eps]{ The projected distribution
of cluster candidates in the LCDCS, subdivided by right ascension. In
all panels, the dashed lines correspond to the declination range of
the survey ($-13\degr<\delta<-11.5\degr$).  The overlaid circles
correspond to the Abell radii for Abell clusters within the field of
view of the survey.
\label{fig:skydist}}
\end{inlinefigure}

\noindent systematic bias in our detection method.  For this sample,
which spans a wide mass range (see Figure \ref{fig:lxtxsig} in \S
\ref{subsec-mass}), our observations have resulted in successful
detection on 16/17 clusters \citep{gon2000thesis}. The undetected
cluster, PDCS 8, has a published 3-$\sigma$ upper limit on the X-ray
luminosity, $L_X=9.8\times10^{43} h_{50}^{-2}$ erg s$^{-1}$ (0.4-2
keV, $\Omega_0=1$) for an estimated redshift $z$=0.6 \citep{hol97}.
We thus anticipate that we are not systematically missing a
significant fraction of the cluster population due to our detection
method, but caution that such a bias may exist.

To quantify the completeness and scatter, we insert the clusters
throughout the entire survey region.  The inserted data consist of
140$\arcsec\times$140$\arcsec$ (200$\times$200 pixel) image sections
centered on the clusters.  Insertion points are chosen at random, but
with the criteria that the position be in the overlap region of the
survey and not lie beneath a mask. No pre-selection is made based upon
extinction or proximity to prominent cirrus features, and so 16\% of
the inserted clusters are rejected on these grounds. These rejected
clusters are not considered in computing the completeness, as they are
already accounted for in the determination of the effective survey
area. We use different random locations for each candidate, and each
of the 12 candidates is inserted into the survey images 264 times (12
insertions per candidate in each of the 22 mosaics).  To avoid
crowding effects, only 72 clusters are inserted into a mosaic for each
run of the simulations.

We track the detection rate and the rate at which a candidate will be
included in the statistical catalog. In addition to determining the
completeness, we also use this data to quantify the rms uncertainty in
$\Sigma_{obs}$.  We find that the standard deviation is
$\sigma_{\Sigma}$=1.6$\times10^{-3}$ counts s$^{-1}$ arcsec$^{-2}$,
independent of $\Sigma_{obs}$, which corresponds to roughly 25\%
uncertainty at the faint limit of the statistical catalog. This
scatter defines the limiting precision with which we 
\begin{inlinefigure}
\plotone{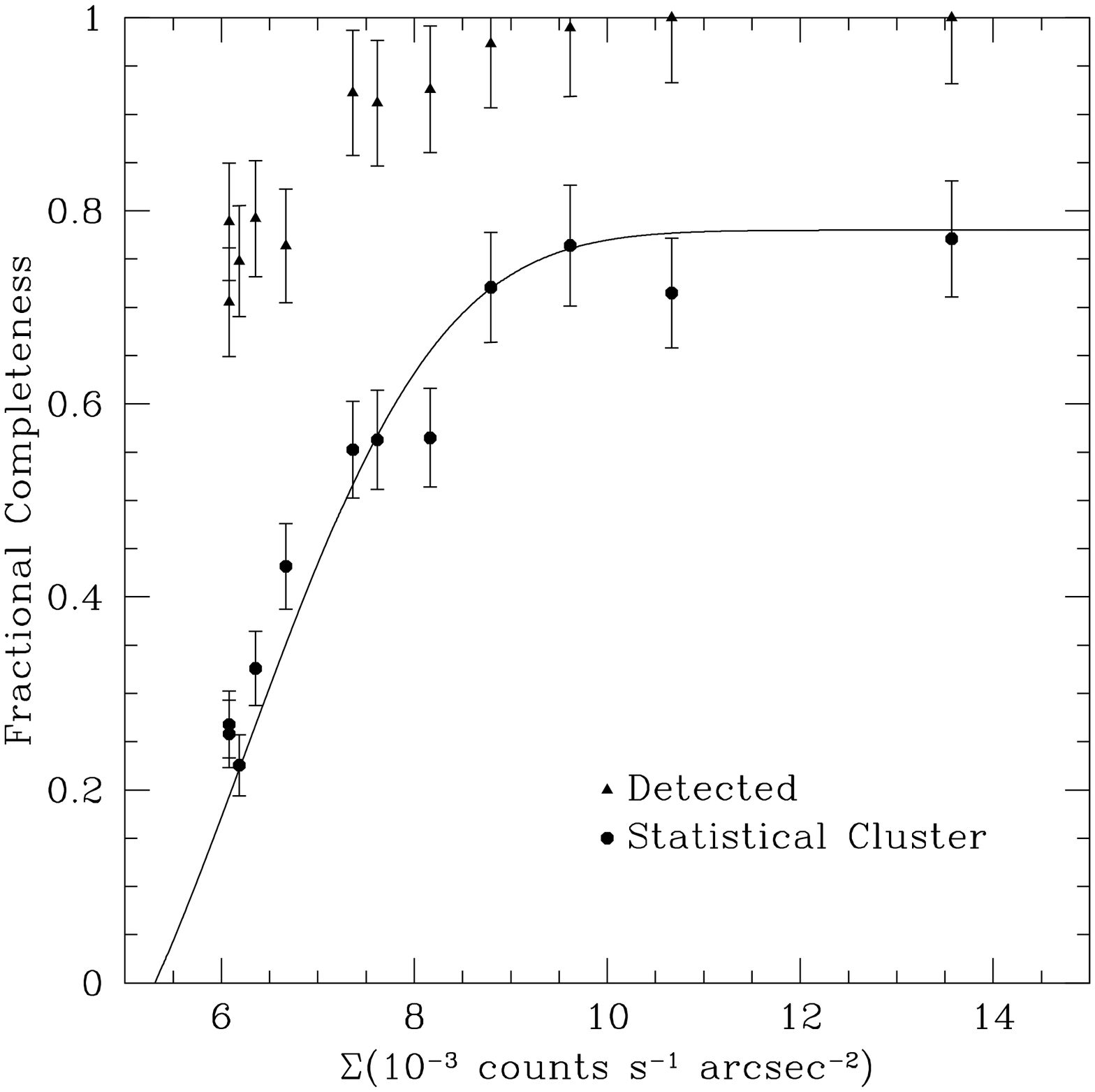}  
\figcaption[f15.eps]{ Fractional completeness of the catalog as
a function of surface brightness. Triangles indicate the probability that a
cluster is detected, assuming that it does not lie directly beneath a mask.
Circles indicate the probability that a cluster is included in the statistical
catalog. The solid curve is an analytic model for the completeness of
the statistical catalog (see text).
\label{fig:completeness}}
\end{inlinefigure}

\noindent can estimate
velocity dispersions for candidates in the sample (see \S
\ref{subsec-mass}).{\footnote{Intrinsic scatter in $\Sigma_{cor}$ is
also a concern; however, the scatter in the $\sigma-\Sigma_{cor}$
relation (see Figure \ref{fig:lxtxsig}$a$) is consistent with arising
purely from our observational uncertainty.}}

 The results of these simulations can be seen in Figure
\ref{fig:completeness} and are listed in Table \ref{table:completeness}.  
The detection rate is near 100\% for
$\Sigma_{obs}$$>$8.5$\times10^{-3}$ \csa. Of these bright candidates,
78\% qualify for inclusion in the statistical catalog. The majority of
those that fail to be included in the catalog are rejected due to
proximity to either a star or a mask, which together eliminate
$\sim$20\% of candidates. Proximity to bright galaxies eliminates most
of the remaining detections. Assuming that the incompleteness at low
surface brightness is the result of Gaussian uncertainty in
$\Sigma_{obs}$, the expected completeness of the statistical catalog
can be modelled as
\begin{equation}
F(\Sigma_{obs})=F_0\times\frac{\mathrm{erf}((\Sigma_{obs}-\Sigma_{lim})/\sqrt{2}\sigma_\Sigma)
+\mathrm{erf}(\Sigma_{obs}/\sqrt{2}\sigma_\Sigma)}{1+\mathrm{erf}(\Sigma_{obs}/\sqrt{2}\sigma_\Sigma)},
\end{equation} 
where $F$ is the fractional completeness, and $F_0$ is the maximum fractional
completeness at high $\Sigma_{obs}$. This model is overlaid in
Figure \ref{fig:completeness} with  
$\Sigma_{lim}$=6.25$\times10^{-3}$ \csa, 
$\sigma_\Sigma$=1.6$\times10^{-3}$ \csa,
and $F_0$=0.78.

\subsection{Contamination Rate}
\label{subsec-contamination}

We utilize follow-up imaging to assess the false detection rate in the
LCDCS. While this approach provides the most direct means of
quantifying the contamination, care must be taken to insure that the
derived result is not biased. The imaging that we have can be divided
into two sets, both of which are described in detail in
\citet{nel2001a}. The first set is deep, multi-color imaging obtained
in 1996 and 1997 with the Las Campanas 1m and 2.5m telescopes. From
these runs, most targets have in excess of 1 hour of imaging on the 1m
(or 20 minutes on the 2.5m) in both $V$ and $I$.  These data are
sufficiently deep to determine whether a candidate is a cluster;
however, some of the targets observed were included specifically
because they looked promising in the drift-scan imaging. As a result,
an assessment of the contamination based upon these data may {\it
underestimate} the true contamination rate.  The second set of data
consists of shallower, $I$-band imaging obtained with the same
telescopes in 1998. For this run care was taken to avoid inducing a
similar selection bias, and the candidate clusters that were observed
are representative of the sample as a whole. Because this data set is
shallow, some true clusters may be incorrectly identified as spurious,
and so an assessment based upon this data may {\it overestimate} the
true contamination rate.

For each candidate we compute the number density of galaxies within a
100 $h^{-1}$ kpc aperture centered on the surface brightness
fluctuation ($\Omega_0$=0.3, $\Omega_{\Lambda}$=0.7; see \S
\ref{subsec-redshifts} for an explanation of the redshift estimates).
We classify targets as clusters if the number density of galaxies in
this aperture exceeds the background level by 2-$\sigma$. Furthermore,
we test the probability that random number density fluctuations in our
fields will exceed this threshold by computing the density contrast
for a complementary set of random locations.  For the deep,
multi-color data set we find that 23/32 (72\%) candidates and 3/32
(9\%) random fields exceed this threshold. We therefore derive a net
contamination rate of 31$\pm$9\% (i.e
$n_{cl}/N=(n-n_{random})/(N-n_{random})$, where $N$ is the total
number of targets, $n$ is the number of $>$2-$\sigma$ detections, and
$n_{cl}$ and $n_{random}$ are the number of cluster and random field
detections).  For the shallower data set we find that 42/57 (74\%) of
candidates and 4/57 (7\%) of random fields exceed this threshold,
yielding a net contamination rate of 28$\pm$7\%.  Because we expect
that the two data sets provide lower and upper bounds on the
contamination rate, the concurrence of these two estimates is
encouraging.  Combining the data sets, our estimate of the
contamination rate is 29$\pm$5\%, with the quoted error bar reflecting
purely Poisson uncertainty.  We also intend to use additional deep,
multicolor observations for contiguous subfields, including data from
the Deep Lens Survey,{\footnote{http://dls.bell-labs.com/}} to provide
an improved constraint in the near future.

Finally, the fractional contamination increases with estimate
redshift, which is a consequence of the method we used to estimate
redshifts.  As described in \S \ref{subsec-redshifts}, we estimate
redshifts for cluster candidates using the BCG magnitude-redshift
relation. For false detections (which by definition lack a BCG), the
galaxy identified as the ``BCG" will be a random field galaxy along
the line of sight. As a result, the distribution of redshifts assigned
to false detections is determined by the magnitude distribution of
field galaxies.  A detailed discussion of the redshift dependence of
the contamination is provided in \citet{gon2000thesis}.

\begin{inlinefigure}
\plotone{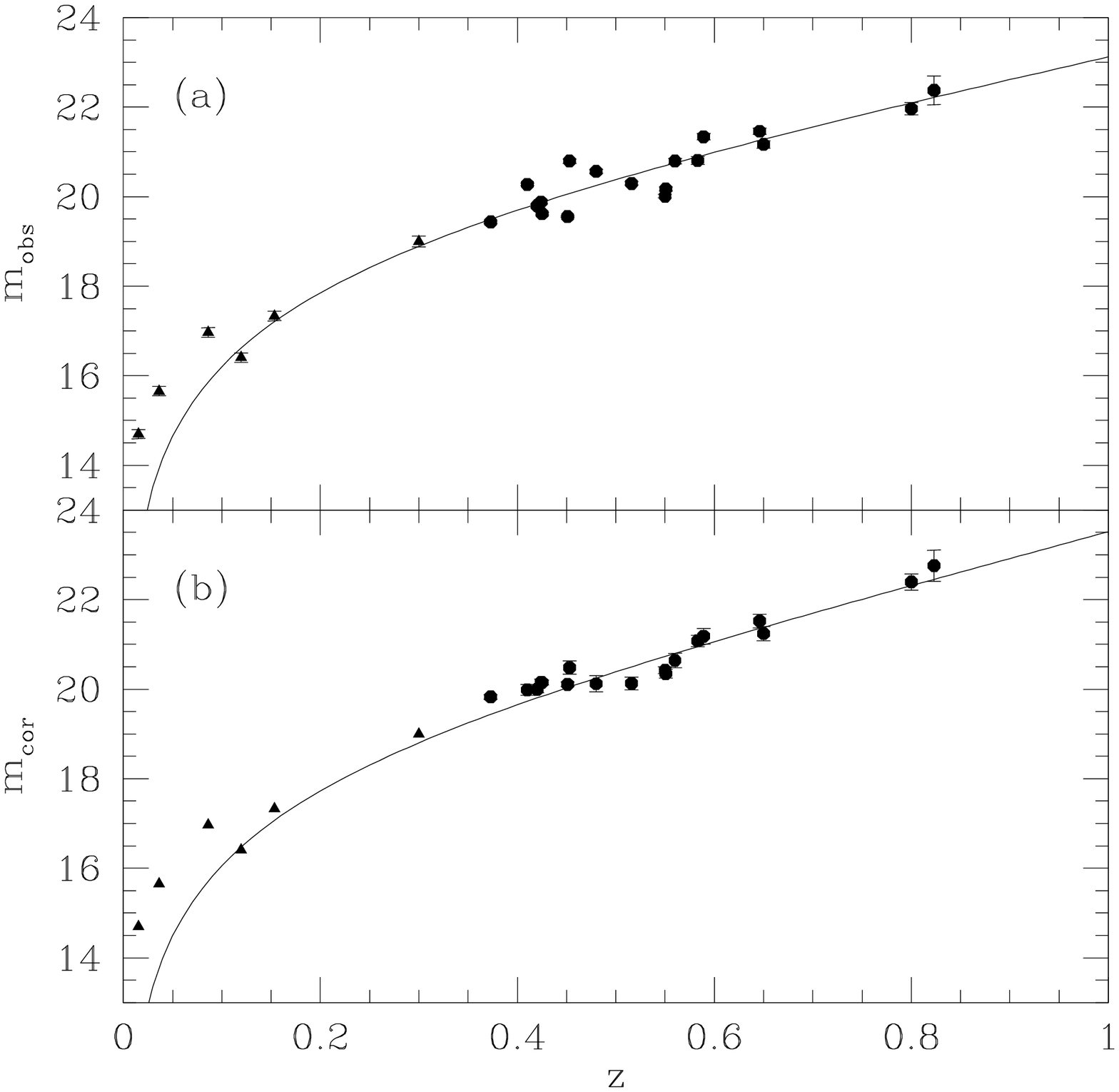} \figcaption[f16.eps]{ (a) BCG magnitude-redshift
relation with no correction for $\Sigma_{cor}$. The solid line is an
unweighted least squares fit to the data of the function given in
Equation \ref{eq:magz}. (b) BCG magnitude-redshift relation including
a $\Sigma_{cor}$ correction. The solid line is an unweighted least
squares fit to the data of the function given in equation (2). For
both plots, only the data points at $z>$0.35 (circles) are used to
calibrate the relation. The lower redshift clusters (triangles), which
are known clusters that lie in our survey region, lack measured values
of $\Sigma_{cor}$. Further, for clusters at $z$$\la$0.1, the
photometric aperture is significantly smaller than the BCG, and so is
a poor estimate of the total luminosity. \label{fig:magz}}
\end{inlinefigure}

\subsection{Redshifts}
\label{subsec-redshifts}

To calculate an empirical redshift estimate, we obtained LRIS
\citep{oke95} spectra at Keck for a subset of 8 LCDCS clusters and 11
clusters from the northern hemisphere pilot survey
\citep{dal95,zar97}. These data were used in conjunction with data for
11 additional previously known clusters to calibrate several
photometric redshift estimators (see Nelson et al. 2000a).  We find
that the most efficient means of estimating redshifts directly from
the survey data is via the brightest cluster galaxy magnitude-redshift
relation.  Locally, brightest cluster galaxies have long been known to
be good standard candles, with dispersions $\sim$0.3 magnitudes
\citep{hum56,san72a,san72b}.  Subsequent work by a number of authors
has found that the scatter remains remarkably small out to at least
z$\sim$1 \citep{san88,ara93,ara98,col98}.

For our calibration data we find comparably small scatter.  We
identify brightest cluster galaxies in an automated fashion -- we
search for the brightest galaxy within a 15$\arcsec$ radii of the peak
and centroid of the surface brightness fluctuation from which a
cluster is detected, and define this galaxy to be the BCG.  This
automated definition is designed to make the catalog easily
reproducible, while minimizing redshift errors due to foreground
contamination or failure to identify the BCG within the chosen search
radius. Because we are detecting the cores of centrally condensed
objects with our technique, there is a high 
\begin{inlinefigure}
\plotone{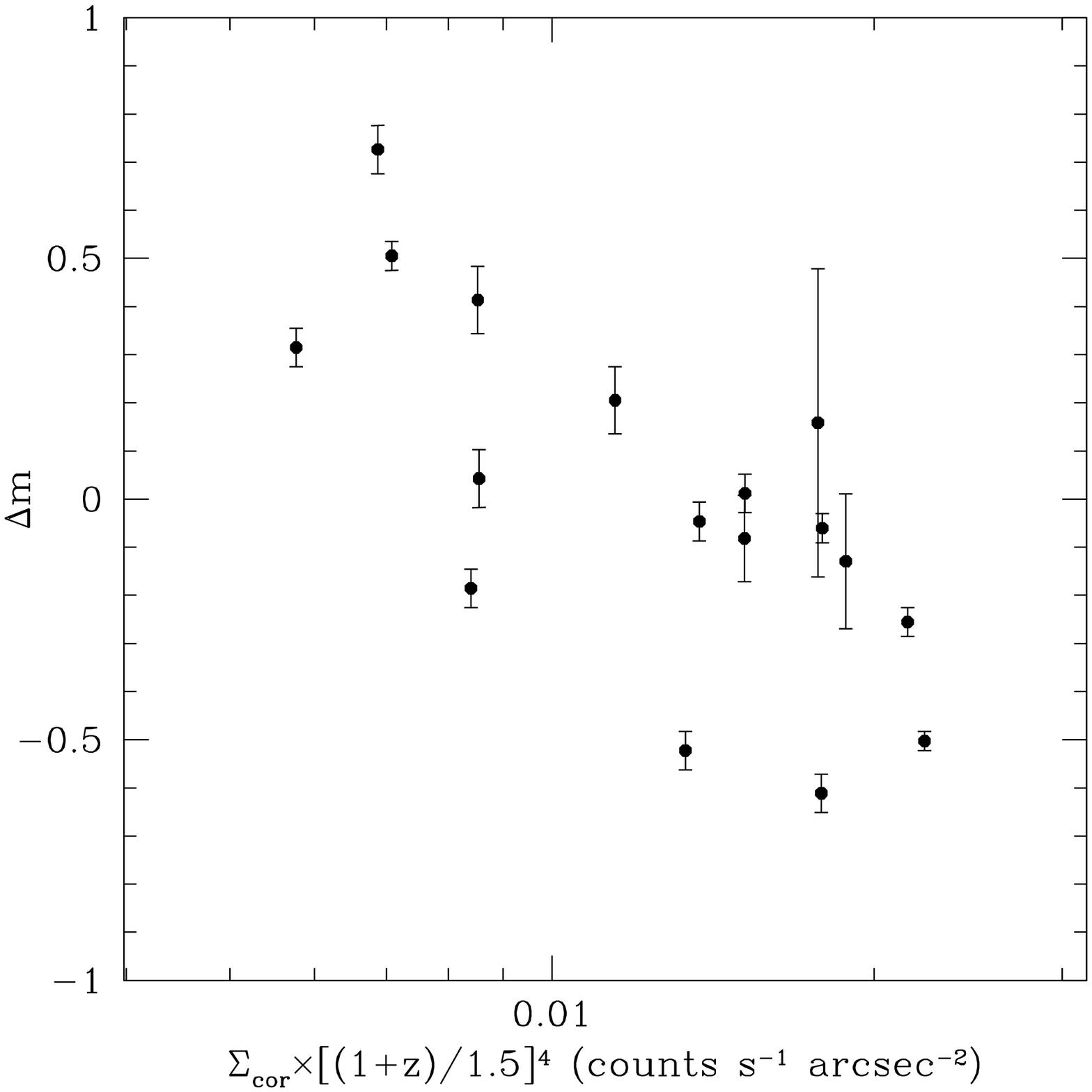} \figcaption[f17.eps]{ BCG magnitude residuals from
Figure \ref{fig:magz}$a$ as a function of $\Sigma_{cor}$, with $\Delta
m\equiv m_{observed} - m_{fit}$.  Error bars correspond to the
photometric uncertainty associated with the observed BCG magnitudes.
\label{fig:magzresidual}}
\end{inlinefigure}

\vskip 1.2in
\noindent probability that the BCG
will be coincident with the surface brightness fluctuation. Further,
if we do miss the BCG, the second ranked galaxy is typically $\sim$0.5
mag fainter \citep{nel2001b}, which corresponds to a redshift error
$\delta z\sim0.1$.  Of the 18 clusters used to calibrate this relation
(the northern clusters were not used because they were observed in a
different filter), in no instance was the galaxy identified as the BCG
more than 0.5 mag fainter than predicted by the final, best-fit model.

We find that the magnitude of the brightest cluster galaxy is an
effective redshift indicator.  Photometry is performed using
SExtractor v2.1.6 \citep{ber96}, and 5$\arcsec$ aperture magnitudes
are used.  All magnitudes are extinction corrected using the maps of
\citet{schl98}.  Figure \ref{fig:magz}$a$ shows the BCG
magnitude-redshift relation for our sample, with the line being a best
fit to the data of the function:
\begin{equation}
m=M+5 \log D_L + B z^{7/4},
\label{eqn-mdist}
\end{equation}                                    
where $M$ is the absolute magnitude and $D_L$ is the luminosity
distance.  The last term is an analytic approximation to the $W$-band
$E+k$ correction, which is derived using the 1995 Bruzual \& Charlot
models \citep{bru93,cha96}.  For the luminosity distance we fix
$\Omega_0$=0.3 and $\Omega_\Lambda$=0; however, precise choice of
cosmology is unimportant due to degeneracy between this term and the
evolution term. The scatter in this relation is $\sigma_m$=0.36 mag
for the calibration data.

To reduce this scatter we test for second-order correlations.
Motivated by an observed correlation between $L_X$ and $M_{BCG}$
\citep{hud97} and our observed correlation between $T_X$ and
$\Sigma_{cor}$ (see \S \ref{subsec-mass} and \citet{gon2000thesis}),
we test for a correlation between $\Sigma_{cor}$ and $m_{BCG}$ (Figure
\ref{fig:magzresidual}). We find that
\begin{equation}
m_{BCG}\propto - \log \Sigma_{cor}(1+z)^4.
\end{equation}
To incorporate this correction, which reduces the scatter to
$\sigma_m$=0.26, we replace $m$ with 
\begin{eqnarray}
m_{corrected}&\equiv& m + A \log \frac{\Sigma_{cor}\ \ \ (1+z)^4}{10^{-2}(1+.5)^4}.
\label{eq:magz} 
\end{eqnarray}
in Equation \ref{eqn-mdist}. Figure \ref{fig:magz}$b$ shows $m_{corrected}$ as
a function of redshift. Overlaid is the unweighted least-squares, which yields
\begin{equation}
m_{corrected} = (-21.3\pm0.15) +5\log D + (1.7\pm0.4) z^{7/4},
\end{equation}
and $A$=1.5$\pm$0.3.  This is the equation that we use to estimate
redshifts for candidate clusters.  For the redshift range of the
LCDCS, a magnitude dispersion of 0.26 mag corresponds to a redshift
uncertainty of 10-11\%. This constitutes a lower limit on the scatter
in our redshift estimates, as there are several caveats to our method.

A first concern is that estimated redshifts beyond $z\sim$0.85 are
based upon extrapolation beyond the redshift range probed by the
calibration sample, and for these clusters the magnitude of the BCG is
very near the detection threshold of our survey data. Consequently, in
Table \ref{table:statistical} 
we simply list the redshift as
$z$$>$0.85 for these clusters. In addition, we caution that the size
of the calibration sample is small, and so our estimate of $\sigma_m$
is subject to small number statistics.

Finally, we expect that the redshift dispersion for the catalog as a
whole is larger than the dispersion for the calibration data due to
foreground contamination and occasional miscentering.  We assess the
robustness of our redshift estimates using the simulations described
in \ref{subsec-completeness}.  These simulations reproduce all sources
of uncertainty in the estimated redshifts (e.g. foreground
contamination, miscentering, scatter in $\Sigma_{obs}$) except for
intrinsic BCG magnitude dispersion.  Failure to identify the BCG due
to miscentering turns out to be a minor issue for the clusters
reinserted into the survey data. We find that for these clusters
redshifts are underestimated by $\Delta z$$>$0.1 less than 2\% of the
time, and only 1\% of the time will a cluster incorrectly be assigned
a redshift $z$$>$0.75.{\footnote{However, if there are systems in
which the BCG is offset from the cluster core by hundreds of kpc, as
suggested by \citet{pos95}, we will fail to identify these BCGs.  In
such cases, we can expect to underestimate the redshift by $\sim$20\%
if we instead identify a second-ranked galaxy (assuming that it is
$\sim$0.5 mag fainter).}} The largest bias in the estimated redshifts
is due to foreground contamination. In Figure \ref{fig:redshiftbias},
this can be seen as a tail towards low redshift.  While negligible at
$z$$<$0.5, by $z$=0.8 foreground contamination leads the redshift to
be underestimated by $\Delta z$$>$0.1 35\% of the time.  In addition,
dispersion in $\Sigma_{obs}$ induces a Gaussian uncertainty in the
estimated redshift of magnitude $\sigma_z$=0.02.  Combined with the
intrinsic dispersion in BCG magnitude, these factors lead us to expect
an rms redshift uncertainty of $\sim$13\% for low-redshift clusters in
the final catalog, rising to $\sim$20\% by $z$$\sim$0.8.  We 
\begin{inlinefigure}
\plotone{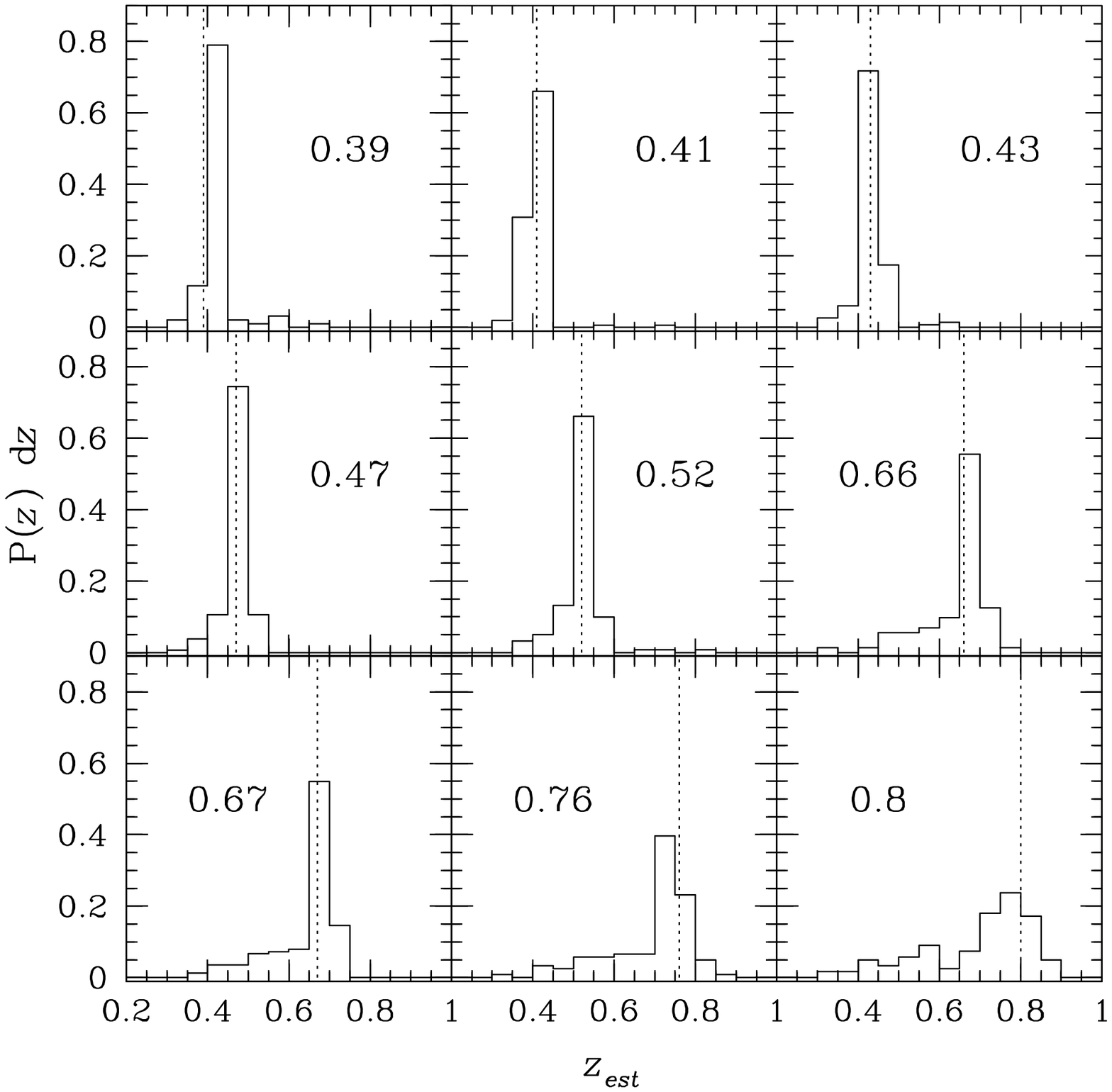} \figcaption[f18.eps]{ Simulated distribution of
estimated redshifts for the nine clusters with
$\Sigma_{obs}$$>$6.25$\times10^{-3}$ \csa. The vertical dashed line is
the original estimated redshift, the value of which is given in each
panel, while the histogram shows the distribution of derived $z_{est}$
when a cluster is randomly reinserted into the survey data. The most
notable systematic bias is a tendency to underestimate the redshifts
of distant systems due to foreground
contamination.\label{fig:redshiftbias}}
\end{inlinefigure}

\noindent emphasize though that this scatter is non-Gaussian due to the 
impact of foreground contamination (i.e. the scatter at $z$=0.8 is 
dominated by a small subset of clusters that have their redshift greatly
underestimated).

\subsection{Velocity Dispersions and X-ray Temperatures}
\label{subsec-mass}

\begin{figure*}
\epsscale{1.00} \plotone{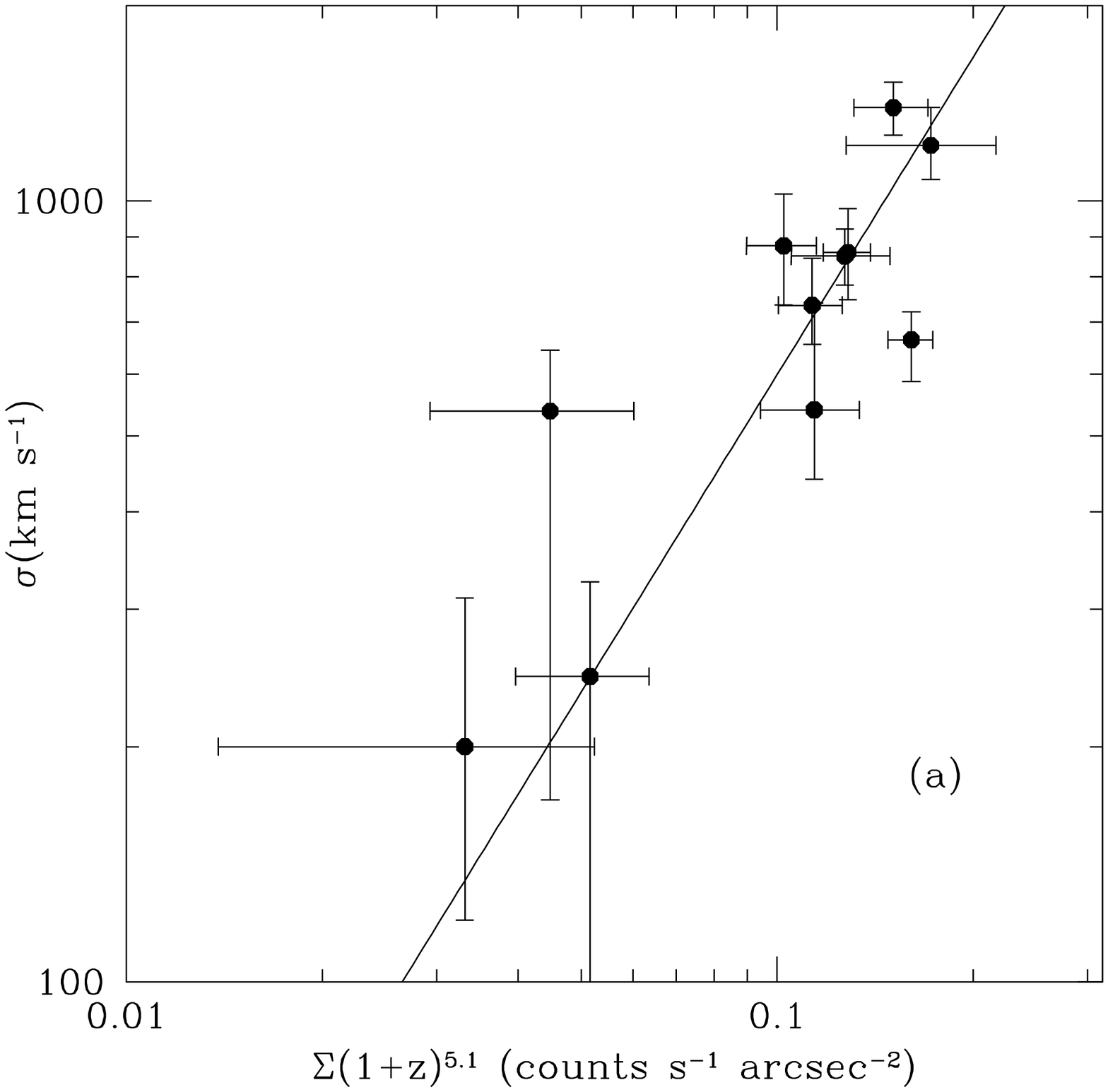} \plotone{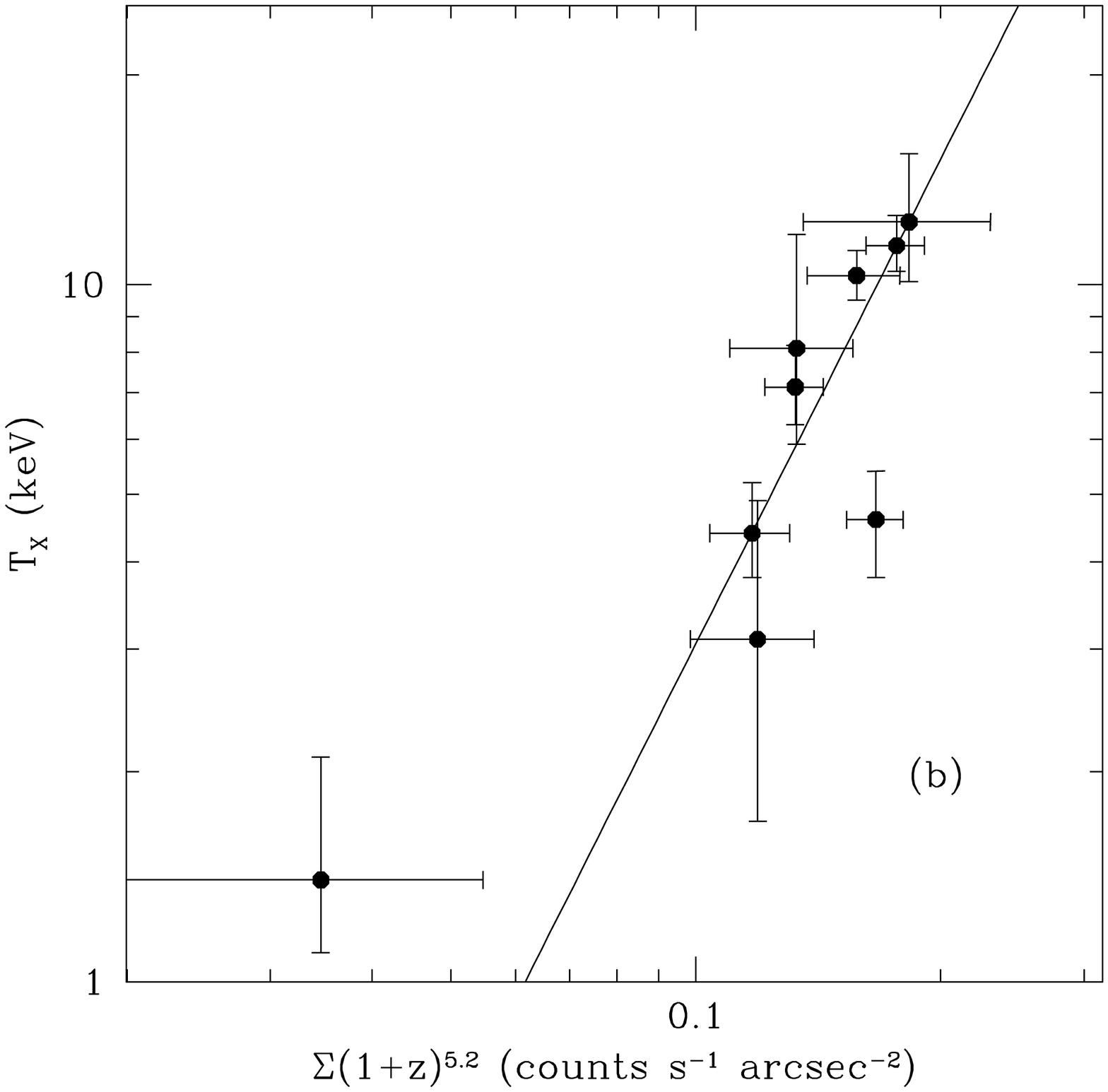} \epsscale{1}
\figcaption[vdsig.eps]{ (a) Velocity dispersion, $\sigma$, as a
function of optical surface brightness, $\Sigma_{cor}$.  The solid
line is the weighted fit from Equation \ref{eqn:vdsig}. For the data
points, $\Sigma_{cor}$ has been multiplied by $(1+z)^{\eta}$ to
eliminate its dependence upon redshift.  (b) Cluster temperature,
$T_X$, vs. $\Sigma_{cor}$. The solid line is the weighted fit from
Equation \ref{eqn:txsig}. For both figures we caution that, due to
small number statistics, the intrinsic scatter in these relations may
be larger the scatter seen in our calibration sample.
\label{fig:lxtxsig}}                        
\end{figure*}

Estimating cluster masses from the survey data is more challenging
than estimating redshifts. To develop a proxy for mass that can be
measured directly from the survey data, we utilize survey quality
drift-scan images of known, X-ray luminous galaxy clusters at
z$>$0.35. As discussed in detail by \citet{gon2000thesis} and shown in
Figure \ref{fig:lxtxsig}, we find $\Sigma_{cor}$ is strongly
correlated with velocity dispersion (linear correlation coefficient
$r$=0.82), X-ray temperature, and X-ray luminosity.  A fit to the
$\sigma$-$\Sigma_{cor}$ data yields
\begin{equation}
\log \sigma=(2.65\pm0.06)+(1.35\pm0.26)\log\left[\frac{\Sigma_{cor}(1+z)^\eta}{10^{-2}(1+.5)^\eta}\right]
\label{eqn:vdsig}
\end{equation}
with $\eta$=5.1$^{+0.39}_{-0.49}$ for $\Omega_0$=0.3 and
$\Lambda=0$.{\footnote {To determine the best fit we minimize the
absolute deviation because this approach is more robust to outliers
than a least-squares fitting procedure (see e.g.  \citet{pre92}).}}
Use of the functional form $(1+z)^\eta$ for the redshift dependence is
based on simulations in which we take detected low-redshift clusters,
artificially move them to higher redshifts (including both
evolutionary and cosmological effects), and then redetect them.  We
find that E+k corrections modify $\eta$ relatively to pure
cosmological dimming ($\eta$=4), but that the functional 

\noindent form remains a good approximation \citep{gon2000thesis}.  The 
$T_X$-$\Sigma_{cor}$ data yield
\begin{equation}
\log T_X=(0.29^{+0.33}_{-0.20})+(2.3\pm1) \log\left[\frac{\Sigma_{cor}(1+z)^\eta}{10^{-2}(1+.5)^\eta}\right]
\label{eqn:txsig}
\end{equation}
with $\eta$=5.2$^{+1.05}_{-0.55}$. The parameter uncertainties for
this fit are significantly larger than for the velocity dispersion
data because of greater uncertainty in $T_X$ than $\sigma$ coupled
with a lack of data for low-mass systems.

To test for consistency, we derive the corresponding $\sigma-T_X$
relation at $z$=0.5 and compare with the results of \citet{xue2000}
for a larger, low-redshift sample. We find $\sigma$=$10^{2.48\pm0.19}
T^{0.59\pm0.28}$, which is consistent with their relation,
$\sigma$=$10^{2.51\pm0.01}T^{0.61\pm0.01}$.  In Figure
\ref{fig:masslimit} we plot the $\sigma$ and $T_X$ corresponding to
the surface brightness limit of the statistical catalog as a function
of redshift for the mass range probed by our calibration
sample. Because \citet{xue2000} note that low-mass groups may obey a
different scaling relation than more massive systems, we refrain from
extrapolating to lower mass. At the lowest redshifts we probe down to
the level of poor groups, while by $z$=0.8 we are only able to detect
very massive systems.{\footnote{Of course, scatter in $\Sigma_{cor}$
will lead to the inclusion of some less massive systems.}}

Finally, one concern with this relation is that it is based upon a
sample of non-LCDCS clusters and so may not probe the range of systems
seen in the LCDCS. In particular, concern has been expressed that, due
to the small size of the smoothing kernel, some of the high-redshift
candidates in the LCDCS may actually be compact groups rather than
massive clusters.  The LCDCS may indeed contain some compact groups;
however, it likely does not contain many of these systems.  Three
factors hinder their detection. First, compact groups tend to lack
central, dominant galaxies, whose halos contribute to the surface
brightness signal. For example, \citet{hic82} found that half of the
first-ranked galaxies in his sample were spirals. Second, compact
groups have small Hubble core radii ($r_c$$\sim$20 $h^{-1}$ kpc,
\citealt{rib98}), which means that at $z$$\ga$0.5 these groups are a
factor of 3.5-4 smaller than the smoothing kernel. This filter
mismatch decreases the observed surface brightness by about 25\% for a
Hubble profile (Figure \ref{fig:filterimpact}).  Third, while compact
groups have high surface densities, much of the total luminosity is
contained in the few most luminous galaxies.  Removal of any of these
galaxies during processing thus has a large impact upon the observed
surface brightness.

\section{Discussion}
\label{sec-discussion}

In this paper we present the Las Campanas Distant Cluster Survey. Our
primary result is a statistical catalog of 1073 cluster candidates at
z$\ga$0.3 drawn from an effective area of 69 square degrees. We also
include a supplementary catalog of 112 candidates that, although they
fail one or more of the automated selection criteria, are strong
cluster candidates. These catalogs together comprise the largest
existing sample of high-redshift clusters, containing roughly three
times more systems than the recently published EIS catalog (302
clusters at 0.2$\la$$z$$\la$1.3).  Even after accounting for an
estimated contamination rate of 30\%, this sample still contains more
candidates at $z$$>$0.3 than all existing published cluster catalogs
combined.  Further, we provide redshift estimates for all candidates,
and also a means of estimating velocity dispersions, enabling
extraction of interesting subsamples for galaxy and cluster evolution
studies.

\begin{inlinefigure}
\plotone{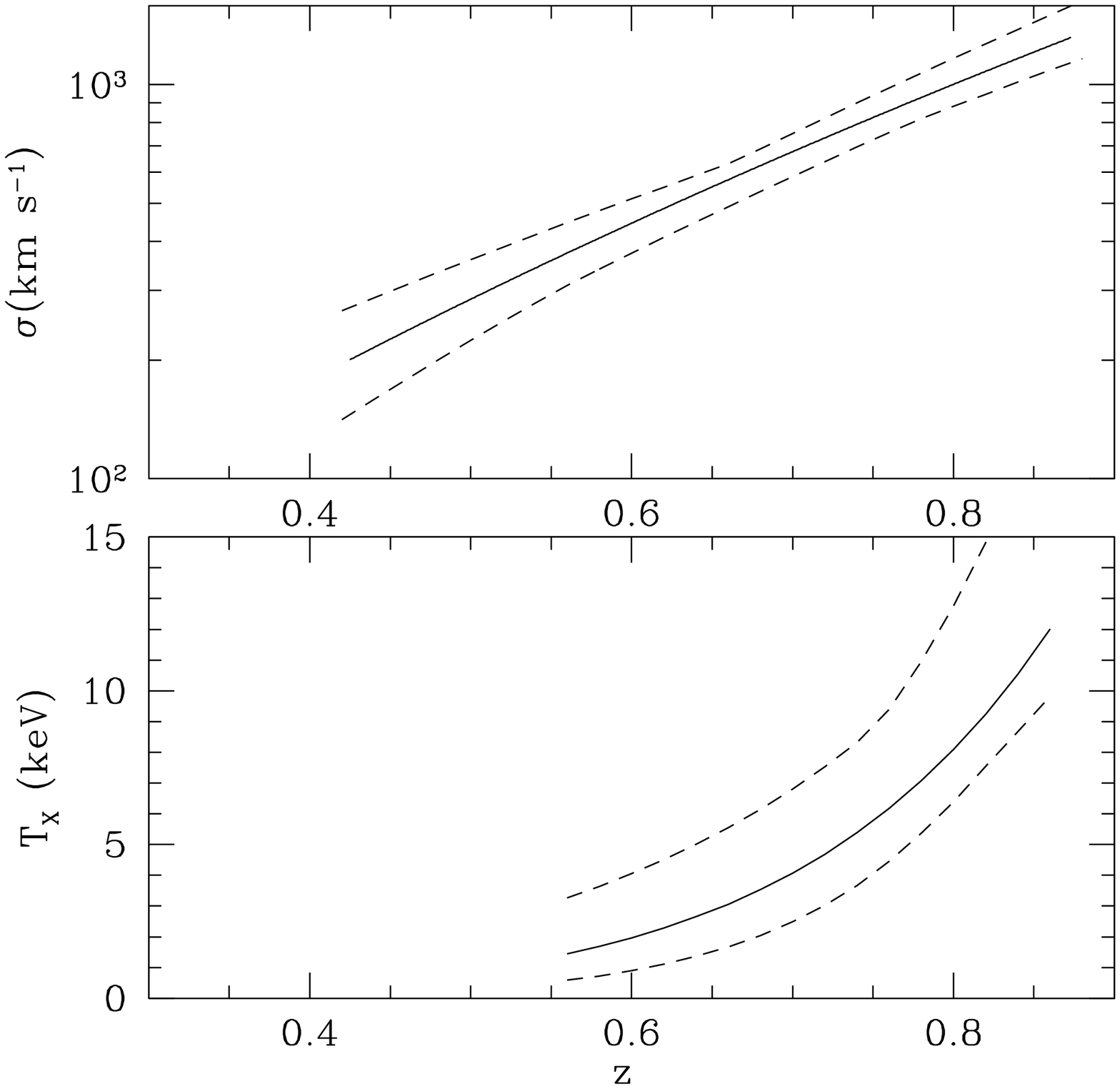} \figcaption[figlim.eps]{ Solid lines are the
limiting velocity dispersions and temperatures of the LCDCS as a
function of redshift, derived from the fits given in Equations
\ref{eqn:vdsig} and \ref{eqn:txsig} with an assumed mean extinction of
E($B$-$V$)=0.05 for the survey. The dashed lines correspond to
1-$\sigma$ uncertainties in the best fit relations. Only the $\sigma$
and $T_X$ regimes covered by the calibration data are plotted.
\label{fig:masslimit}}
\end{inlinefigure}

Finally, to a large degree this survey is an experiment in the
feasibility of constructing large catalogs of clusters using surface
brightness fluctuations. In the next few years, traditional optical
surveys that utilize information about the individual cluster galaxies
will provide catalogs of comparable size to the LCDCS \citep{gla00},
however with the deep imaging required by these surveys it is unlikely
that they will provide another order of magnitude improvement beyond
the LCDCS in the near future. 

\noindent In contrast, drift-scan imaging of a
much larger fraction of the sky will exist within the next few
years. Consequently, the greatest value of this survey is perhaps not
the final catalog, but rather the demonstration that this technique
will work in an automated fashion with large data sets.

With this in mind, we consider modifications and improvements upon the
reduction procedure and data presented in this paper that will enable
construction of better, larger catalogs in the future. What are the
primary problems and limitations of the LCDCS? What features would
significantly improve the usefulness of this sample? The most obvious
answers are greater redshift depth, greater angular area, and lower
contamination -- especially at the highest redshifts.  Additionally,
to push to higher redshift than the LCDCS, improved redshift estimates
are required since at the limit of our survey foreground contamination
is beginning to become a serious issue.

To achieve these improvements, several fairly straightforward tactics
can be utilized. First, observing in two or more colors, while
increasing the required observing time, would significantly reduce the
contamination rate.  Multi-color information can be used to eliminate
system-related signals, and also to discriminate cluster-induced
\begin{inlinefigure}
\plotone{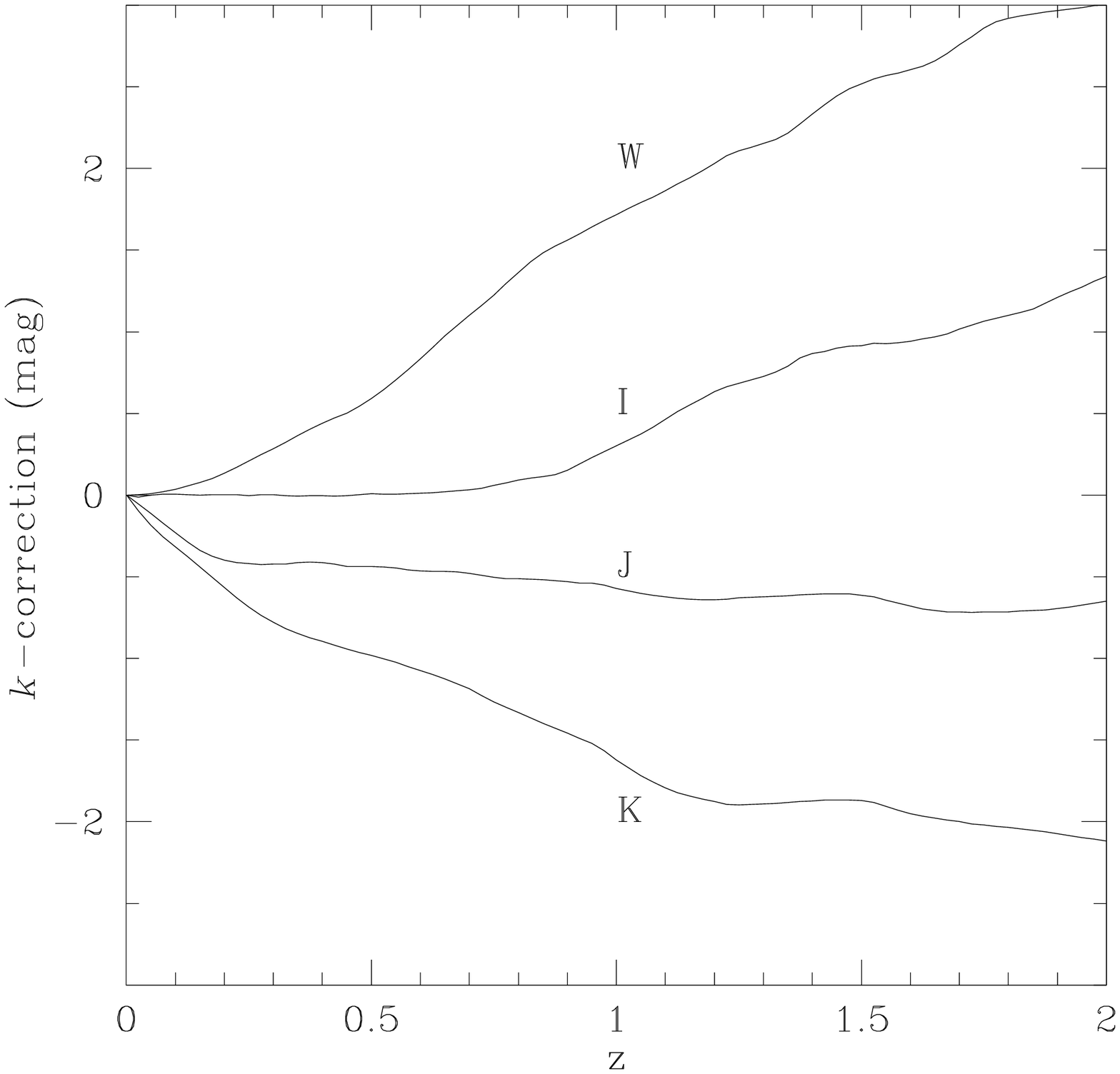} \figcaption[kcor.eps]{ The $k$-correction for a
passively evolving elliptical in the LCDCS $W$-band and several longer
wavelength passbands. Curves are based upon the models of
\citet{cha96}. \label{fig:kcor}}
\end{inlinefigure}

\noindent fluctuations (which should be relatively red) from other sources such
as low surface brightness galaxies and galactic cirrus.  Further,
color information can be used to generate improved photometric
redshift estimates and, by permitting color selection of brightest
cluster galaxies, minimize foreground contamination.

Improving survey depth is more challenging. While the $W$ filter
utilized in this survey was designed to maximize the incident optical
flux, the $k$-correction for a passively evolving elliptical is
large. At $z$=1 the $k$-correction corresponds to an effective dimming
of two magnitudes relative to rest frame, and by a redshift of two the
net dimming is roughly three magnitudes.  To probe to higher redshift,
it is necessary to shift to redder wavelengths.  Figure \ref{fig:kcor}
shows $k$-corrections for a passively evolving elliptical in several
passbands.  $K^\prime$ is the ideal choice (with $J$ or $H$ as the
second color), as the $k$-correction is actually quite negative out to
z$\sim$2. A challenge with using near IR is that the sky level is very
high and rapidly varying, and so additional care must be taken to
avoid spurious detections, Further, unlike CCD's, near IR detectors
read out individual pixel elements independently, which precludes
drift scanning in the fashion employed in the LCDCS.  Still, if these
observational difficulties can be overcome, near IR surface brightness
surveys hold the potential to identify clusters, out to z$\approx$2.

A more conservative but more immediately feasible option is to use
slightly shorter wavelength passbands, such as $I$ and $z$ to conduct
wide-area surveys. Here the $k$-corrections are worse than in
$K^{\prime}$, but large-area drift-scan data are already in
existence. The Sloan data set in particular is ideal for such a
survey.  The Sloan Survey is currently obtaining
$u^{\prime}$-,$g^{\prime}$-, $r^{\prime}$-, $i^{\prime}$-, and
$z^{\prime}$-band imaging data on the Apache Point 2.5m.{\footnote{See
\citet{fuk96} for details on the Sloan filters.}} This data set should
be sufficient to detect clusters out to $z\approx$1.25.{\footnote{For
expected limiting magnitudes, see
http://www.sdss.org/science/tech\_summary.html .}}  Using the
$g^{\prime}$, $r^{\prime}$, and $i^{\prime}$ passbands, a
contamination rate of $\la$10\% should be possible, and BCG magnitudes
can be used to generate redshift estimates with $\sigma_z$$\la$0.10
for all candidates (alternatively, the multiwavelength data
potentially can be used to derive more precise photometric redshifts).
The goal of Sloan is to image $\pi$ steradians, or roughly 10,000
square degrees centered on the north galactic cap.  If this entire
area were analyzed using a technique similar to the one used in the
LCDCS, of order 10$^5$ clusters and groups could be detected at
$z\ga$0.5, including 5000-10000 clusters with $T_X$$>$5 keV.

Several additional minor improvements should also be considered for
implementation in future surveys. An adaptive filter should probably
be employed for smoothing, as this will eliminate assumptions
regarding the characteristic scale of clusters at high
redshift. Alternatively, wavelet analysis could provide a means of
eliminating these assumptions, while also improving removal of large
scale, background variations. Finally, the fractional area masked can
be greatly reduced if PSF models accurate to large radii are used to
remove saturated stars. Unfortunately this procedure proved
impractical for the LCDCS data set, but would have resulted in
recovery of $\sim$30 square degrees.

Detection of clusters via surface brightness fluctuations has great
potential for dramatically increasing the number of known massive
clusters. With this survey we have demonstrated the viability of this
approach, in the process generating a catalog of over 1000 cluster
candidates at z$>$0.3. This sample is currently being used to study
galaxy evolution \citep{nel2001a}, probe large scale structure as
traced by clusters, and constrain cosmological models via evolution in
the cluster number density \citet{gon2000thesis}. We are also working
to quantify the relative detection efficiency of this and other
optical algorithms with the aim of better understanding the selection
biases intrinsic in each approach.  Finally, it is our hope that this
work will serve as the basis for larger, deeper surveys in the future
that will further extend the redshift baseline and permit even more
detailed study of the cluster population.

\begin{acknowledgements}
We thank Rebecca Bernstein for providing the Savitsky-Golay code
employed in bias and sky subtraction and the Carnegie Observatories
for their generous allocation of telescope time for this project. We
also the thank Marc Postman for his thorough analysis and constructive
comments in refereeing this paper and Stefano Andreon for identifying
a mistake in \S 3.  AHG acknowledges support from the National Science
Foundation Graduate Research Fellowship Program, the ARCS Foundation,
and the Harvard-Smithsonian Center for Astrophysics. DZ acknowledges
financial support from National Science Foundation CAREER grant
AST-9733111, and fellowships from the David and Lucile Packard
Foundation and Alfred P. Sloan Foundation. JD acknowledges support
from NASA grant HF-01057.01-94A and GO-07327.01-96A. AEN acknowledges
financial support from a National Science Foundation grant
(AST-9733111) and the University of California Graduate Research
Mentorship Fellowship program.  This research has made use of the
NASA/IPAC Extragalactic Database (NED) which is operated by the Jet
Propulsion Laboratory, California Institute of Technology, under
contract with the National Aeronautics and Space Administration.
\end{acknowledgements}

\begin{appendix}
\section{Projection Effects and Astrometric Alignment}

To accurately align the drift scans and generate mosaics in which
objects are aligned to $\delta\theta\la1\arcsec$ over the entire
length of the scans, two issues must be addressed. First, we must
account for geometrical projection effects. Failure to correct for
projection effects leads to a maximal alignment error of 2$\arcsec$ in
declination. Second, and more importantly, we must transform data
taken along arbitrary Great Circles to a common coordinate system
before alignment is possible. This is particularly critical because a
subset of the data was taken along lines of constant right ascension
due to mechanical difficulties with the Great Circle Camera (GCC), and
so in some cases it is necessary to align scans of this type with
Great Circle scans. In these cases, failure to transform the GCC scans
would result in alignment errors of several arcminutes at the ends of
the scans.  For extensive information on projection effects,
astrometric alignment, and related topics, we refer the reader to
\citet{mcn74}.

Projection effects are typically a two-dimensional issue; however, the
problem is simplified with drift scan data. Let us define an angular
coordinate system ($\tau,\beta$), where $\tau$ represents the
longitudinal angle along the Great Circle traced by the GCC, and
$\beta$ is the declination relation to this circle. Further, define
the projected cartesian coordinate system ($x,y$), with $x$ being the
direction perpendicular to the direction of the scan. For a pointed
observation, the standard gnomic formulae are
\begin{equation}
x=\beta_0+\frac{\tan\beta\cos\beta_0-\sin\beta_0\cos(\tau-\tau_0)}
{\tan\beta\sin\beta_0+cos\beta_0\cos(\tau-\tau_0)}, \label{eq:xgnomorig}
\end{equation}
and,
\begin{equation}
y=\tau_0+\frac{sin(\tau-\tau_0)}
{\tan\beta\sin\beta_0+cos\beta_0\cos(\tau-\tau_0)}, \label{eq:ygnomorig}
\end{equation}
where ($\tau_0,\beta_0$) are the coordinates of the center of the
image.  With a drift scan the situation is slightly different.  Along
the direction of the scan, projection effects are washed out because a
given position on the sky is observed over the full range of the
detector. To see this mathematically, consider equation
\ref{eq:ygnomorig}. Since we are now averaging over a range of $\tau$,
we should integrate over the second term. Further, since all values of
$\tau$ are equivalent there is no longer a central $\tau_0$ and so the
appropriate equation for $y$ is
\begin{equation}
y=\tau+\int_{\tau-\Delta\tau}^{\tau+\Delta\tau}\frac{sin(\tau^\prime-\tau)}
{\tan\beta\sin\beta_0+cos\beta_0\cos(\tau^\prime-\tau)} d\tau^\prime.
\end{equation}
(see Figure \ref{fig:proj1}). Since the centerline of the scan lies along
the Great Circle traced by the scan, $\beta_0=0$, which simplifies this
last equation to
\begin{equation}
y=\tau+\int_{\tau-\Delta\tau}^{\tau+\Delta\tau}\tan(\tau^\prime-\tau)
d\tau^\prime= \tau,
\end{equation}
with the second term disappearing because we are integrating over an
odd function. For $x$, there is no such averaging effect. Still, we
can simplify the projection equation. We have already stated that
$\beta_0=0$. We now also note that for a drift scan the gnomic
projection in $x$ is independent of $\tau$, and so set $\tau_0=\tau$
in equation \ref{eq:xgnomorig}.  Thus, the equation for $x$ simplifies
to
\begin{equation}
x=0+\frac{\tan\beta\cos0-\sin0\cos0} {\tan\beta\sin0+\cos0\cos0}=\tan\beta,
\end{equation}
and so the equations of projection for a drift scan along a Great Circle are
\begin{equation}
y=\tau,
\end{equation}
\begin{equation}
x=\tan\beta.
\end{equation}

The above equations adequately describe the gnomic distortion of the
images, but we are still in a coordinate system based upon the Great
Circle along which the scan was taken. Next, it is necessary to
transform these coordinates to equatorial right ascension and
declination ($\alpha,\delta$). We once again refer the reader to
\citet{mcn74} for a more thorough treatment of spherical coordinate
transformations. Define ($\alpha_G,\delta_G$) to be the coordinates of
the northern pole of the GCC system in equatorial coordinates,
($\alpha_0,\delta_0$) to be the equatorial coordinates of the center
of the scan, and ($\tau_P,\beta_P$) to be the Great Circle coordinates
of the northern pole of the equatorial system. The coordinates of the
pole and scan center are related as
\begin{equation}
(\alpha_G,\beta_G)=
        \begin{cases}
         (\alpha_0+180,90-\delta_0) &   \mathrm{if} \;\;\alpha_0<180,\cr
         (\alpha_0-180,90-\delta_0) &   \mathrm{if} \;\;\alpha_0>180.\cr
        \end{cases}
\end{equation}
With these definitions, application of spherical trigonometry leads to the
relations
\begin{equation}
\sin \delta = \sin \delta_G \sin \beta + \cos\delta_G\cos\beta\cos(\tau_P-\tau),
\mathrm{\textbf{\ \ \ \ \ \ \ \ \ \ \ Cosine\ Law} }
\end{equation}
\begin{equation}
\cos\delta\sin(\alpha-\alpha_G) = \cos\beta\sin(\tau_P-\tau),
\mathrm{\textbf{\ \ \ \ \ \ \ \ \ \ \ \ \ \ \ \ \ Sine\ Law} }
\end{equation}
and
\begin{equation}
\cos\delta\cos(\alpha-\alpha_G)=\sin\beta\cos\delta_G -
\cos\beta\sin\delta_G\cos(\tau_P-\tau).
\mathrm{\textbf{\ \ Analog\ Formula} }
\end{equation}
It is these three equations that we use to transform the survey data.
Note that although we have three equations relating the coordinates,
these do not over-constrain the problem.  The third equation is
required to remove sign ambiguities.  Subsequent to these
transformations, the cartesian coordinates of the survey correspond to
a 1-to-1 linear map of right ascension and declination, with a plate
scale of 0$\farcs$7 pixel$^{-1}$.

\begin{inlinefigure}
\plotone{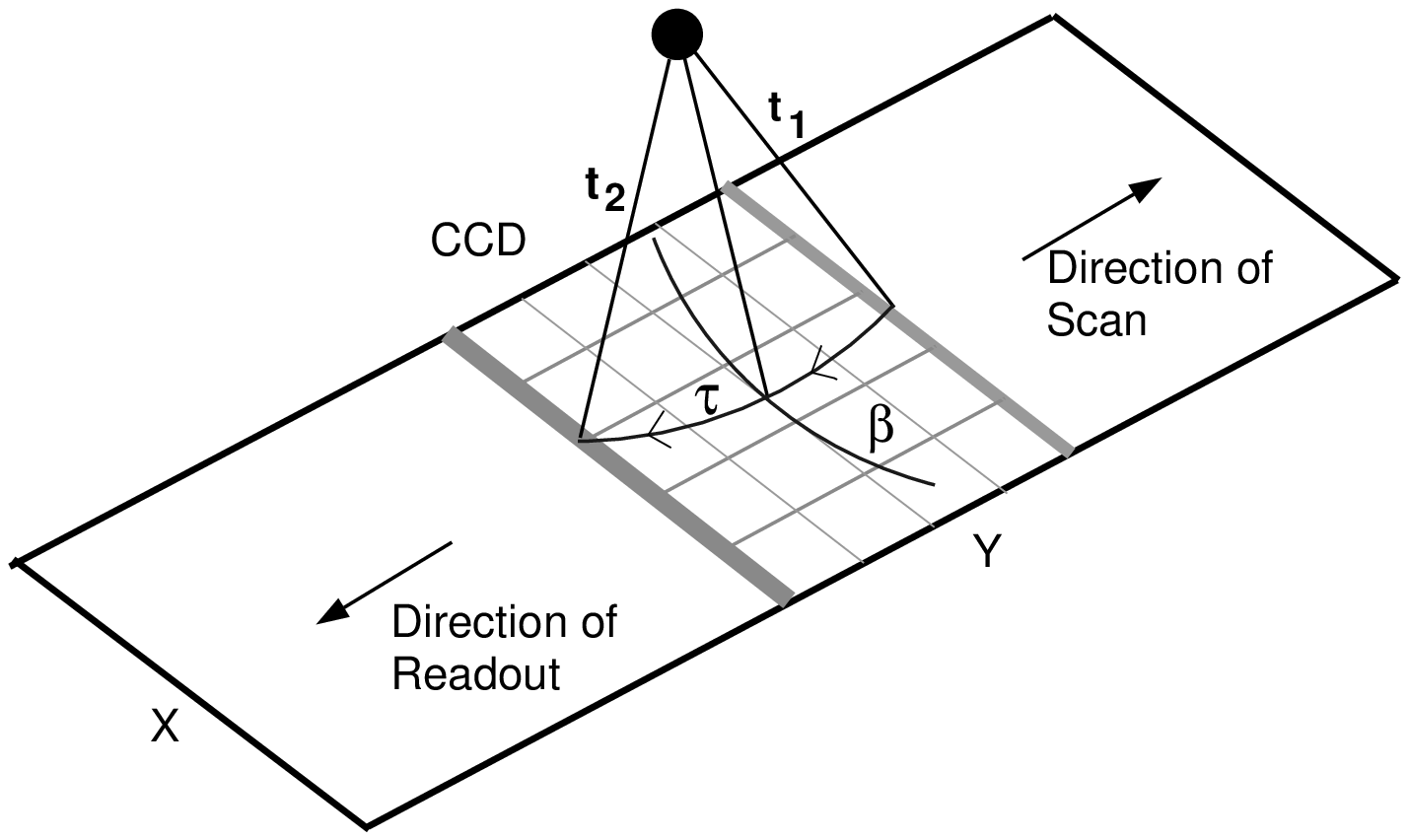} \figcaption[ccd.ps]{Illustration of projection
effects for drift scanning.  Projection effects are washed out along
the direction of readout as the signal is averaged across the entire
ccd (with time increasing from $t_1$ to $t_2$).  Perpendicular to the
readout, projection of $\beta$ onto $x$ must be corrected.
\label{fig:proj1}}
\end{inlinefigure}              
\end{appendix}

\clearpage
\input{table1.tex} 
\input{table2.tex} 
\input{table3.tex} 

\end{document}

%% file: table1.tex
\begin{deluxetable}{rccrccrrrrcl}
\tablecaption{Statistical Catalog}
\tablewidth{0pt}
\setlength{\tabcolsep}{0.04in}

\tablehead{
\colhead {ID} & \multicolumn{3}{c}{$\alpha$(J2000)}  &
\multicolumn{3}{c}{$\delta$(J2000)} &
\colhead {$z_{est}$} & \colhead{$\Sigma_{obs}$} &
\colhead{$\Sigma_{cor}$} & \colhead{E($B$-$V$)} & \colhead {Notes}  \\
\colhead{} & \colhead{h} & \colhead{m} & \colhead{s} &
\colhead{$\degr$} & \colhead{$\arcmin$} &\colhead{$\arcsec$} &
\colhead{} & \colhead{} & \colhead{} & \colhead{} & \colhead{}
}
\startdata
0001   & 10&01&07.4   & -12&26&24   & 0.80  & 6.46  & 7.84  & 0.070  & \\
0002   & 10&02&01.5   & -12&45&35   & 0.43  & 6.79  & 8.05  & 0.062  & $z_{Keck}$=0.52\\
0003   & 10&02&17.8   & -12&23&15   & 0.37  & 6.38  & 7.82  & 0.074  & \\
0004   & 10&02&17.9   & -12&50&10   & 0.59  & 7.25  & 8.62  & 0.062  & \\
0005   & 10&02&27.1   & -12&47&13   & 0.57  & 7.62  & 9.17  & 0.067  & \\
0006   & 10&02&39.3   & -11&39&32   & 0.41  & 6.41  & 7.81  & 0.071  & \\
0007   & 10&02&43.6   & -11&35&16   & 0.50  & 7.06  & 8.66  & 0.074  & \\
0008   & 10&02&46.6   & -12&38&42   & $>$0.85  & 6.57  & 8.00  & 0.071  & \\
\enddata
\tablecomments {Units for $\Sigma_{obs}$ and $\Sigma_{cor}$ are 10$^{-3}$ counts s$^{-1}$
arcsec$^{-2}$.
(1) RX J1347.5 at $z$=0.451 \citep{sch95}
}
\label{table:statistical}
\end{deluxetable}

%% file: table2.tex
\begin{deluxetable}{rccrccrrrrccl}
\tablecaption{Supplemental Catalog}
\tablewidth{0pt}
\setlength{\tabcolsep}{0.04in}

\tablehead{
\colhead{} & \multicolumn{3}{c}{$\alpha$(J2000)}
&\multicolumn{3}{c}{$\delta$(J2000)} & \colhead{} & \colhead{}
& \colhead{} & \colhead{} & \colhead{Failed} &\colhead{} \\
\colhead {ID} & \colhead{h} & \colhead{m} & \colhead{s} & \colhead{$\degr$} &
\colhead{$\arcmin$} &\colhead{$\arcsec$} &
\colhead {$z_{est}$}
&\colhead{$\Sigma_{obs}$} & \colhead{$\Sigma_{cor}$} &
\colhead{E($B$-$V$)} & \colhead{Criteria$^{a}$} &\colhead {Notes}
}
\startdata
S001   & 10&06&18.8   & -12&58&13   & 0.47  &  9.85  & 11.80  & 0.066  & S & \\
S002   & 10&06&25.8   & -12&40&48   & 0.60  &  7.77  &  9.05  & 0.055  & S & \\
S003   & 10&15&38.4   & -12&43&52   & 0.39  & 11.40  & 14.00  & 0.075  & M & \\
S004   & 10&23&11.1   & -12&33&47   & 0.29  & 13.40  & 16.60  & 0.079  & $z$,G & \\
S005   & 10&24&09.8   & -11&48&08   & 0.29  & 21.10  & 24.30  & 0.051  & $z$,M & \\
S006   & 10&27&21.4   & -11&47&48   & 0.50  &  9.75  & 11.20  & 0.051  & M & \\
S007   & 10&28&33.7   & -11&59&29   & 0.35  & 12.50  & 14.60  & 0.057  & G & \\
S008   & 10&31&27.9   & -12&35&20   & 0.56  &  7.63  &  9.17  & 0.067  & S & \\
\enddata
\tablecomments{($a$) Automated criteria that the detection failed to satisfy.
The following designations are used: E - extinction E(\bv)$\ge$0.10; $z$ - $z$$<$0.30; 
LSB - identified as a low surface brightness galaxy; S - flagged as likely
spurious; M - partially masked; G - identified as a galaxy because
m$_T$$<$18.75.
}
\label{table:supplemental}
\end{deluxetable}

%% file: table3.tex
\begin{deluxetable}{ccccc}
\tablecaption{Detection and Classification Results}
\tablewidth{0pt}
\tablehead{
\colhead{$\Sigma_{obs}$} & \colhead{} & \colhead{Fraction} & \colhead{Statistical} \\
\colhead {(10$^{-3}$ csa)} & \colhead {$z_{est}$}
&\colhead{Detected} & \colhead{Cluster Fraction}
}

\startdata
6.08 & 0.52 & 0.82 & 0.26 \\ 
6.08 & 0.52 & 0.73 & 0.27 \\ 
6.19 & 0.66 & 0.77 & 0.23 \\ 
6.35 & 0.66 & 0.81 & 0.33 \\ 
6.67 & 0.39 & 0.78 & 0.43 \\ 
7.36 & 0.52 & 0.93 & 0.55 \\ 
7.62 & 0.76 & 0.92 & 0.56 \\ 
8.16 & 0.80 & 0.93 & 0.56 \\ 
8.79 & 0.47 & 0.98 & 0.72 \\ 
9.61 & 0.43 & 0.99 & 0.76 \\ 
10.67 & 0.41 & 1.00 & 0.71 \\ 
13.57& 0.67 & 1.00 & 0.77 \\ 
\enddata
\label{table:completeness}
\end{deluxetable}